\documentclass[11pt]{article}

\usepackage[T1]{fontenc}
\usepackage[utf8]{inputenc}
\usepackage[a4paper,margin=2.5cm]{geometry}
\usepackage{amsmath,amssymb}
\usepackage{graphicx}
\usepackage{xcolor}
\usepackage{caption}
\usepackage{pythonhighlight}
\usepackage{pdfpages}
\usepackage{pdflscape}
\usepackage{url}

\usepackage{natbib}
\setcitestyle{authoryear,round,aysep={},yysep={,},notesep={, }}

\usepackage[colorlinks=true,linkcolor=blue,citecolor=blue,urlcolor=blue,
            breaklinks=true]{hyperref}

\graphicspath{{img/}}
\begin{document}

\title{\bf Exoplanet Detection Techniques: Radial Velocity}

\author{
  Rafael Luque\thanks{Corresponding author: \texttt{rluque@iaa.es}}\\[2pt]
  \small Instituto de Astrof\'isica de Andaluc\'ia (IAA-CSIC), Granada, Spain\\[6pt]
  Matthew R. Standing\\[2pt]
  \small European Space Agency (ESA), European Space Astronomy Centre (ESAC), Madrid, Spain
}
\date{}

\maketitle

\begin{abstract}
\noindent
The radial velocity (RV) technique enabled the discovery of the first
exoplanets orbiting Sun-like stars. Since then, improvements in precision and
sensitivity have enabled the detection of smaller planets, bringing us closer
to the goal of discovering Earth analogs in the coming years.
In this chapter, we provide an overview of the radial velocity method. We begin
with a detailed description of the technique, including how precise radial
velocities are measured and what limitations must be addressed to achieve
reliable detection of small signals.
We examine the unique challenges associated with the detection of circumbinary
planets and outline the developments that have, for the first time, made their
detection feasible using the radial velocity method.
With over 2,000 planets detected to date, the radial velocity method remains
one of the most reliable techniques, offering valuable insights into the
demographics of exoplanets, especially at long orbital distances. As an example
of its applications, we provide an analysis and interpretation of three
publicly available radial velocity datasets: one from a single star, one for a
circumbinary planet system, and another from a single star with transiting
planets.
Furthermore, we compile information on the spectrographs currently and
soon-to-be available worldwide for radial velocity observations and a list of
open-source tools designed for their analysis. In the Supplementary Material,
we include a sample telescope proposal for obtaining radial velocity data,
along with practical tips on writing a competitive proposal to aid early-career
researchers entering the field.
This chapter serves as a comprehensive guide for understanding the capabilities
and applications of the radial velocity method, equipping readers with the
knowledge and resources to address their own scientific questions using this
technique.

\end{abstract}

\tableofcontents
\bigskip

\section{Introduction}\label{sec:intro}

\begin{figure}
	\centering
	\includegraphics[width=\textwidth]{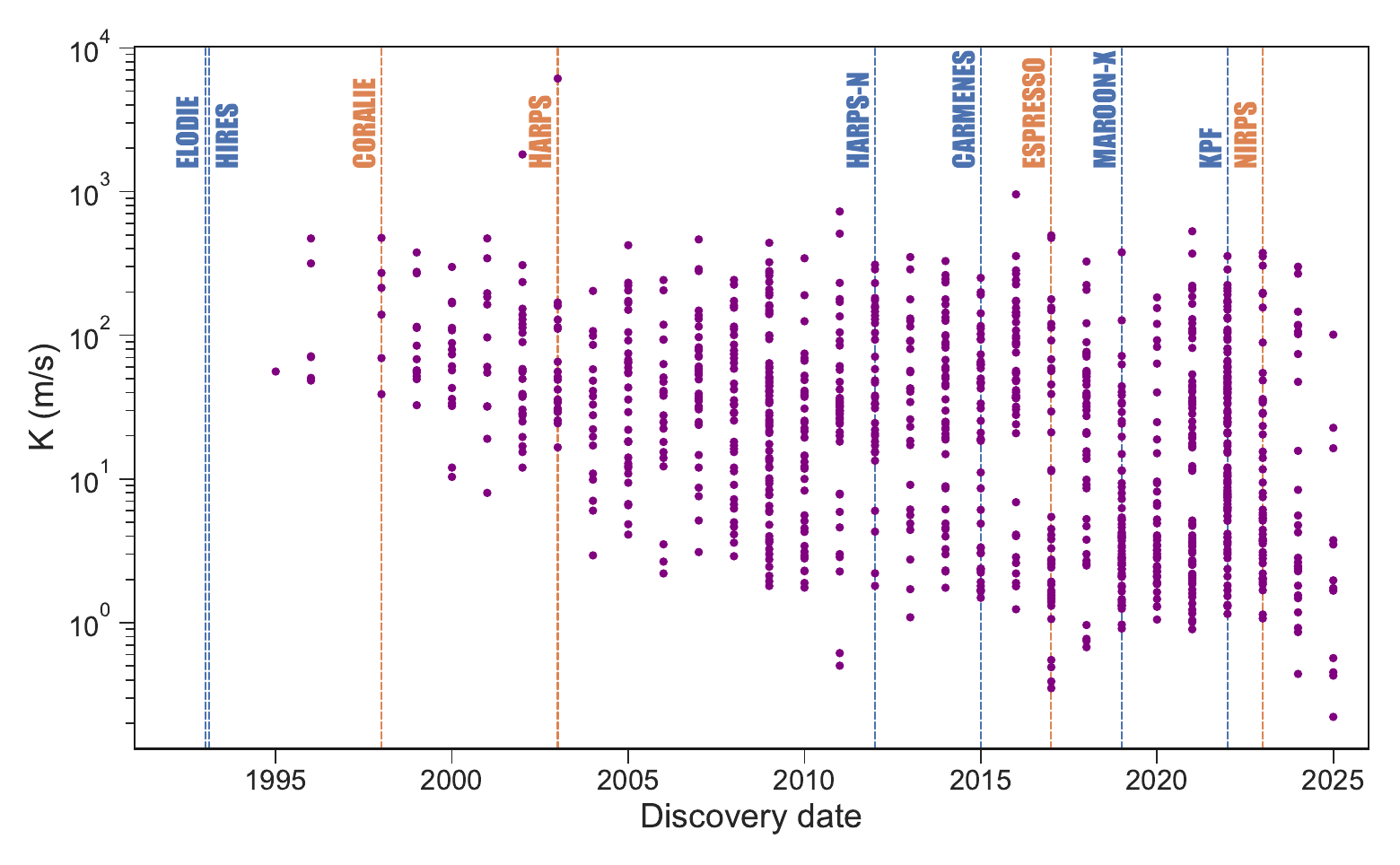}
	\caption{The radial velocity semi-amplitude ($K$) of exoplanets discovered with the radial velocity technique as a function of discovery date. Vertical dashed lines mark the first light of some of the most relevant instruments used for precise radial velocity studies. The color relates to their location (northern and southern hemispheres in blue and orange, respectively). Data obtained from the NASA Exoplanet Archive (\url{https://exoplanetarchive.ipac.caltech.edu}) in April 2025.}
	\label{fig:NEA_YearRV}
\end{figure}

At the time of writing, nearly six thousand exoplanets --- planets that orbit stars different than the Sun --- have been discovered. This chapter focuses on the use of the radial velocity technique for the detection and characterization of exoplanets. This method enabled the first discoveries and drove the remarkable growth of this field of astrophysics over the past four decades. The radial velocity technique produces exoplanet discoveries at approximately 100 per year. Figure~\ref{fig:NEA_YearRV} shows the progress over the past 30 years, where not only the number of discoveries has increased, but also technological developments have allowed signals as small as 22\,cm\,s$^{-1}$ to be detected \citep{Basant2025}. The plot also reveals an important feature: planets with radial velocity semi-amplitudes of about 1\,m\,s$^{-1}$ or larger are routinely detected, but smaller signals are much more challenging. As we will discuss in subsequent sections, variability from the stars themselves is the ultimate hurdle to reaching the necessary precision to constrain the occurrence of terrestrial planets in the habitable zone of Sun-like stars.

However, the concept of radial velocity is present not only in the exoplanet field but in all of astronomy. In 1842, Christian Doppler theorized that a wave will change its wavelength depending on the relative motion of the source emitting it and an observer. This ``Doppler effect'' therefore means that an observer will detect the light from a star at a slightly different wavelength than when emitted. This difference is given by the following equation \citep[e.g.,][]{CarrollOstlie1996}:

\begin{equation}\label{eq:doppler_effect}
	\lambda = \lambda_0 \frac{\sqrt{1+\frac{\rm {v_r}}{c}}}{\sqrt{1-\frac{\rm {v_r}}{c}}}
\end{equation}
where $\lambda$ is the detected wavelength of the light, $\lambda_0$ is the wavelength of the emitted light, $\rm {v_r}$ is the velocity of the star with respect to us along our line of sight, and $c$ is the speed of light in a vacuum. Therefore, light is shifted to shorter wavelengths (to the blue end of the spectrum or blueshifted) when its source moves towards us, and to longer wavelengths (to the red end or redshifted) when the source is moving away. 

Doppler originally proposed that this effect could be used to explain the color of binary stars. Despite being incorrect, the effect was easily demonstrated for sound waves and applied to electromagnetic waves (mainly by Fizau). Among the applications of the Doppler effect in astronomy we highlight the demonstration that Saturn's rings were not solid, but made of small particles \citep{Keeler1895ApJ.....1..416K}; the demonstration of the expansion of the Universe \citep{Hubble1929PNAS...15..168H}; the first indirect evidence of the existence of dark matter via the flat rotation curves of galaxies \citep{Rubin1978ApJ...225L.107R, FaberGallagher1979ARA&A..17..135F}; and the discovery of the first exoplanet orbiting a Sun-like star \citep{Mayor1995} envisioned by Otto Struve in his (now) seminal paper from 1952 \citep{Struve1952}.

In this chapter, we will cover the following aspects of the radial velocity technique. Section~\ref{sec:Meth_desc} describes the method, including how to measure precise radial velocities, the influence of stellar activity on radial velocity measurements, and the challenges of detecting planets orbiting binary stars using this technique. Section~\ref{sec:demographics} summarizes some of the most important population trends revealed by radial velocity discoveries, from giant to terrestrial planets. Section~\ref{sec:Analysis} contains examples of the analysis of real radial velocity observations of single and binary stars using publicly available software. The chapter ends with information about the instruments used (in the past, present, and future) to measure precise radial velocities and a compilation of open-access tools for their analysis, modeling, and interpretation. In the Supplementary/Online material (\ref{sec:proposal_example}) we include an example of a successful observing proposal used to discover and characterize a planet in a binary system using radial velocity measurements.

\section{Method description}\label{sec:Meth_desc}

\begin{figure}
	\centering
	\includegraphics[width=\textwidth]{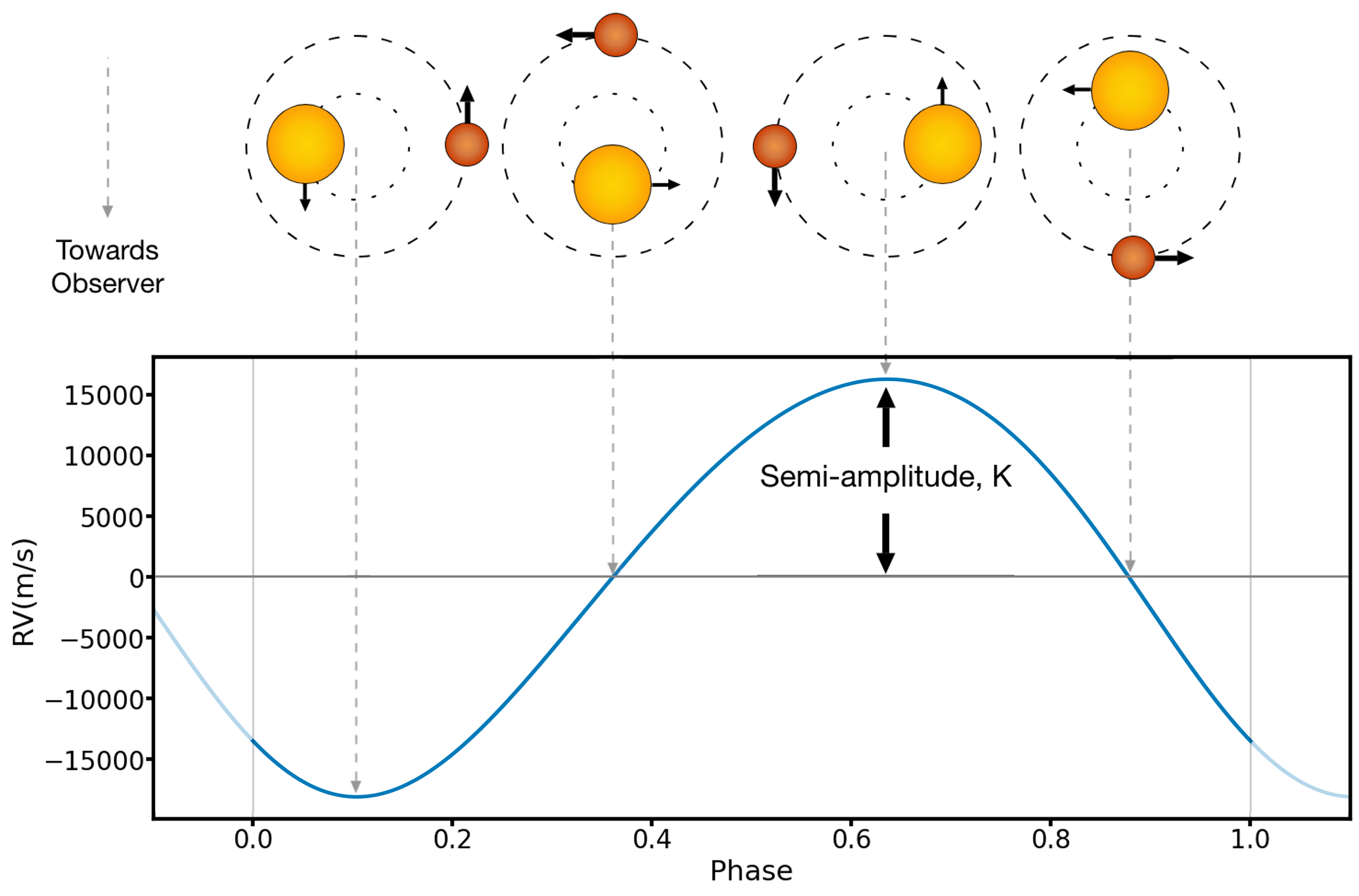}
	\caption{Top: Schematic of a binary star system orbital motion four times through one orbit. Bottom: Phased radial velocity signal of the primary star. The blue line is a Keplerian radial velocity signal. The semi-amplitude $K$ of the binary signal is shown. Adapted from \citet{Standing2022b}.}
	\label{fig:RV_schematic}
\end{figure}

In the context of exoplanet research the radial velocity technique, also known as the Doppler technique, relies on the fact that the presence of a planetary companion with mass $M_{\rm p}$ around a host star with mass $M_\star$ displaces the center-of-mass of that two-body system from the center of the star itself. As a result, a star with a sole companion exhibits a Keplerian orbit around the system's barycenter due to the gravitational pull of the orbiting planet. The stellar motion along the line-of-sight of the observer is known as the \emph{radial velocity}, which is the fundamental measured quantity naming this technique --- although a precise definition requires additional considerations \citep[see][for a full discussion]{LindegrenDravins2003}.  This back-and-forth wobble of the star can be measured using the Doppler effect, thus constraining the properties of the body orbiting it (see Fig.\ref{fig:RV_schematic}). Here, we derive the radial velocity equation following a similar approach as in \citet{Seager2010exop.book.....S}.

\subsection{Radial velocity signature of Keplerian motion}

\begin{figure}
	\centering
	\includegraphics[width=\textwidth]{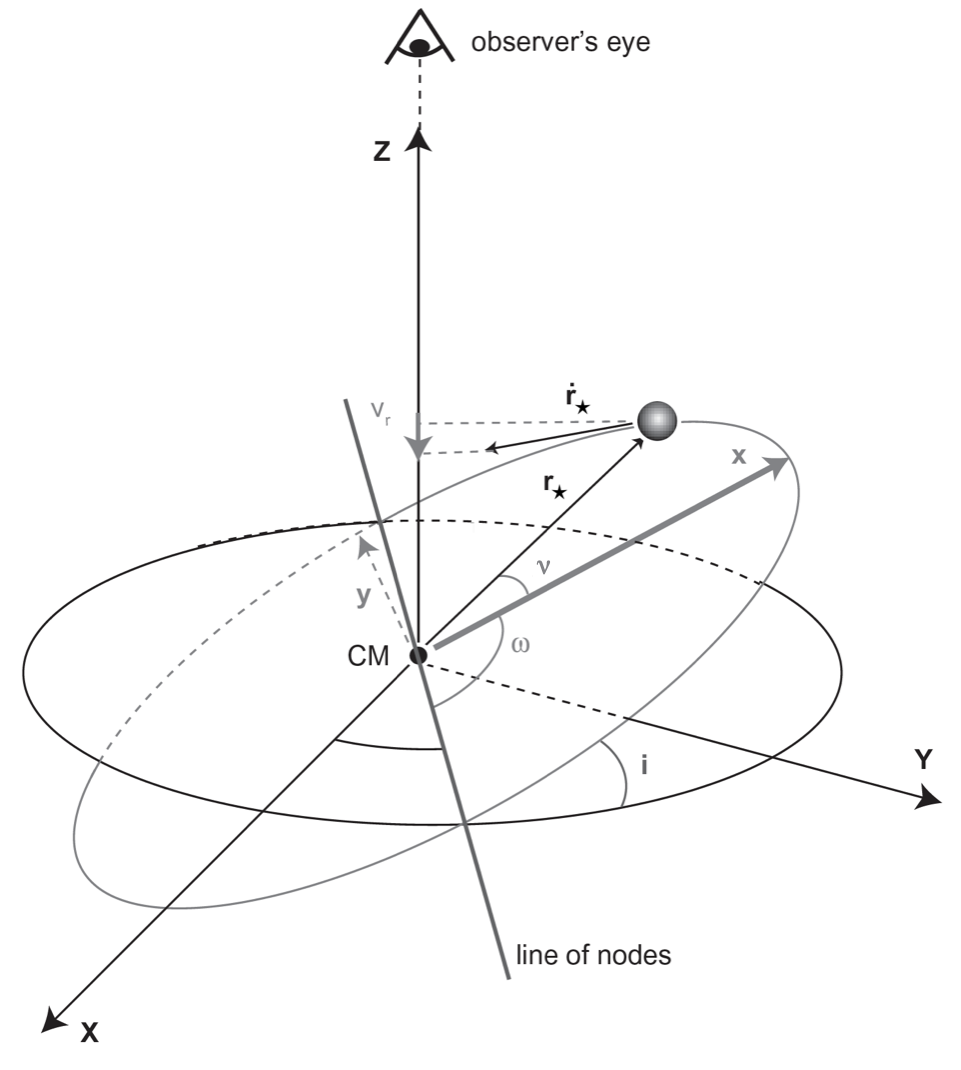}
	\caption{The relationship between the star's velocity around the center of mass, $\bf \dot{r}_\star$, and its radial component along the line of sight, $\rm v_{\rm r}$. Adapted from \citet{MurrayCorreia2010exop.book...15M}.}
	\label{fig:orbital-geometry}
\end{figure}

The equation of the ellipse described by the star around the center of mass is, in polar coordinates,
\begin{equation}\label{eq:rstar}
	{\rm r_\star} = \frac{a_\star\left(1-e^2\right)}{1+e\cos {\rm \nu}}
\end{equation}
where $r_\star$ is the distance of the star from the barycenter, $e$ is the eccentricity, and $\rm \nu$ is the true anomaly, i.e., the angle between the periastron direction and the position on the orbit measured from the center of mass. The semi-major axis of the star around the barycenter, $a_\star$, is related to the orbital semi-major axis $a$ through 
\begin{equation}\label{eq:astar}
	a_\star = \frac{M_{\rm p}}{M_\star + M_{\rm p}} a \quad .
\end{equation}

Since the observable quantity we are interested in is the stellar radial velocity, it is necessary to obtain the relation between the position of the orbit and orbital velocity. In Cartesian coordinates, following the naming convention in Fig.~\ref{fig:orbital-geometry} with the x-axis (in gray, which is different from the X-axis in the reference system of the observer) pointing in the star's periastron direction, the origin at the barycenter, and the z-axis perpendicular to the orbital plane; the position and velocity vectors are given by
\begin{equation} \label{eq:dotr}
	{\bf r_\star} = 
	\begin{pmatrix} 
		{\rm r_\star}\cos {\rm \nu} \\
		{\rm r_\star}\sin {\rm \nu} \\
		0 \\
	\end{pmatrix} \quad , \quad
	\dot{\bf r}_\star = 
	\begin{pmatrix} 
		{\dot{\rm r}_\star}\cos {\rm \nu} - {\rm r_\star}\dot{\rm \nu}\sin {\rm \nu} \\
		{\dot{\rm r}_\star}\sin {\rm \nu} + {\rm r_\star}\dot{\rm \nu}\cos {\rm \nu} \\
		0 \\
	\end{pmatrix} \quad .
\end{equation}

Differentiating Eq.~\ref{eq:rstar} to express the velocity as a function of the true anomaly,
\begin{equation}
	\dot{\rm r}_\star = 
	\frac{a_\star e \left(1-e^2\right)\dot{\rm \nu}\sin {\rm \nu}}{\left( 1+e\cos {\rm \nu}\right) ^2} = 
	\frac{e\,\rm r^2_\star\, \overset{.}{\nu} \sin {\rm \nu}}{a_\star\left(1-e^2\right)} \quad ,
\end{equation}
and replacing it into Eq.~\ref{eq:dotr}, we obtain
\begin{equation} \label{eq:dotr2}
	\dot{\bf r}_\star = \frac{\rm r^2_\star\,\overset{.}{\nu}}{a_\star\left(1-e^2\right)}
	\begin{pmatrix} 
		-\sin {\rm \nu} \\
		e+ \cos {\rm \nu} \\
		0 \\
	\end{pmatrix} = \frac{\rm h_\star}{M_\star a_\star\left(1-e^2\right)}
	\begin{pmatrix} 
		-\sin {\rm \nu} \\
		e+ \cos {\rm \nu} \\
		0 \\
	\end{pmatrix} \quad ,
\end{equation}
where ${\rm h}_\star = M_\star\,{\rm r}^2_\star\,\dot{\rm \nu}$ is the angular momentum of the star, which is a constant of motion. This quantity can be expressed as a function of the ellipse parameters $a$ and $e$ \citep[see][]{MurrayCorreia2010exop.book...15M} as
\begin{equation}\label{eq:hstar}
	{\rm h}_\star = \sqrt{\frac{G\,M^2_\star\,M^4_{\rm p}\,a\,\left(1-e^2\right)}{\left(M_\star + M_{\rm p}\right)^3}} \quad .
\end{equation}

Substituting Eq.~\ref{eq:astar} and \ref{eq:hstar} in Eq.~\ref{eq:dotr2}, we finally obtain
\begin{equation} \label{eq:dotr3}
	\dot{\bf r}_\star = \sqrt{\frac{G\,M^2_{\rm p}}{\left(M_\star + M_{\rm p}\right)}
		\frac{1}{a\left(1-e^2\right)}}
	\begin{pmatrix} 
		-\sin {\rm \nu} \\
		e+ \cos {\rm \nu} \\
		0 \\
	\end{pmatrix} \quad .
\end{equation}

The final step is to project the velocity vector onto the observer's line of sight. The inclination angle $i$ of the system is defined as the angle between the orbital plane and the plane of the sky (i.e., the perpendicular to the line of sight). The argument of periastron $\omega$ is defined as the angle between the line of nodes and the periastron direction (see Fig.\ref{fig:orbital-geometry}). The radial velocity equation is obtained by projecting the velocity vector on the line of sight unit vector $\hat{\bf k}$, with x and y axes in the orbital plane and z perpendicular to them, as
\begin{equation} \label{eq:vstar}
	{\rm v_{\rm r,\star}} \equiv {\rm v_{\rm r}} = \dot{\bf r}_\star \cdot \hat{\bf k} = 
	\dot{\bf r}_\star \cdot 
	\begin{pmatrix} 
		\sin\omega\sin i \\
		\cos\omega\sin i \\
		\cos i \\
	\end{pmatrix} =  
	\sqrt{\frac{G}{\left(M_\star + M_{\rm p}\right)a\left(1-e^2\right)}}
	M_{\rm p}\sin i \left( \cos(\omega + {\rm \nu}) + e\,\cos\omega \right)
	\quad .
\end{equation}

From here the radial velocity semi-amplitude $K = ({\rm v}_{\rm r,max} - {\rm v}_{\rm r,min})/2$ is derived, the fundamental observable quantity relating the radial velocity to the position on the orbit. 
\begin{equation} \label{eq:k}
	K = 
	\sqrt{\frac{G}{\left(1-e^2\right)}}
	M_{\rm p}\sin i \left(M_\star + M_{\rm p}\right)^{-1/2}\,a^{-1/2}
	\quad .
\end{equation}

Replacing the semi-major axis $a$ with the orbital period $P$ using Kepler's third law and expressing this formula in more practical units, we obtain
\begin{equation} \label{eq:k2}
	K = 
	\frac{28.4329\,\mathrm{m\,s^{-1}}}{\sqrt{1-e^2}}
	\frac{M_{\rm p}\sin i}{M_J}
	\left(\frac{M_\star + M_{\rm p}}{M_\odot}\right)^{-2/3}
	\left(\frac{P}{1\,\mathrm{yr}}\right)^{-1/3} \quad .
\end{equation}

\begin{figure}
	\centering
	\includegraphics[width=\textwidth]{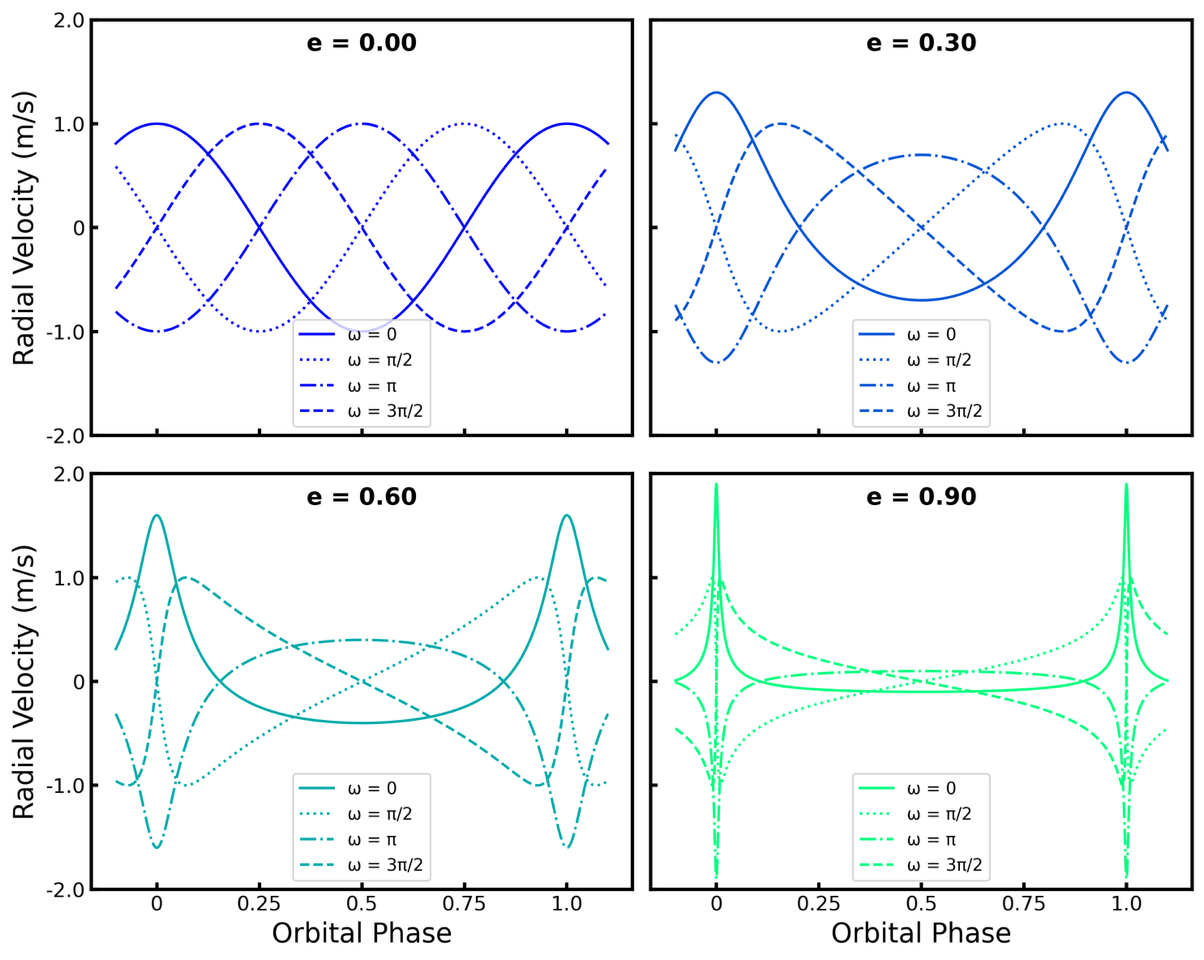}
	\caption{Keplerian radial velocity curves calculated using Eq.~\ref{eq:vstar} when varying orbital eccentricity $e$ and $\omega$. $K$, $P$, and $T_0$ are constant.}
	\label{fig:RV_eccentricity}
\end{figure}

From a radial velocity time series, five orbital parameters can be determined by fitting a Keplerian orbit: $e$, $\omega$, $P$, $K$, and the time of periastron passage $T_0$. Varying the orbital eccentricity $e$ (how circular an orbit is) in Eq.~\ref{eq:vstar} above can yield radial velocity curves with very different shapes, as shown in Fig.\ref{fig:RV_eccentricity}.

Prior knowledge of the stellar mass allows us to determine the mass of the planetary companion from the radial velocities using Eq.~\ref{eq:k2}. However, the radial velocity technique, in the exoplanet case where $M_{\rm p} \ll M_\star$, is only sensitive to the minimum mass $M_{\rm p}\sin i$ of the companion unless there is a priori information from the system's orbital inclination $i$ from transit or astrometric observations (see Chapters~3 and 6, respectively). For planets orbiting binary star systems, known as circumbinary planets, if the binary stars experience eclipses, this reveals the inclination angle of the system \citep{Triaud2013}. The $\sin i$ factor in the companion's mass illustrates the geometric effect of the radial velocity signal when the orbit is edge-on ($i=90\,\mathrm{deg}$) or face-on ($i=0\,\mathrm{deg}$) from the observer's line of sight.

Equation~\ref{eq:k2} applies to any two celestial bodies. The radial velocity (RV) semi-amplitude is higher for more massive companions around low-mass objects at short-period orbits. For a star orbiting a stellar companion with a similar mass, $K$ is on the order of km\,s$^{-1}$. For an Earth-mass planet orbiting a Sun-like star with an orbital period of 1\,yr (a configuration that we will refer to as ``Earth-Sun analog system''), the RV semi-amplitude is only 9\,cm\,s$^{-1}$, comparable with the walking speed of a giant tortoise. Examples of the detection and characterization of planetary systems orbiting single and binary stars using this technique are given in Section~\ref{sec:Analysis}.

\subsubsection*{Resonant double versus eccentric single}
Keplerian radial velocity signals from multiple planets are combined when monitoring the motion of the star. This combination can lead to an incorrect interpretation of the source of the Keplerian signals. For example, two planets in a 2:1 orbital resonance (where the orbital period of the outer planet is close to twice that of the inner planet) can appear as a single eccentric planet in radial velocity data \citep[e.g.][]{Anglada2010, Wittenmyer2013, Boisvert2018, Nagel2019}. This effect appears stronger when the outer circular planet is the more massive of the two \citep{Anglada2010}. \citet{Anglada2010} find that around $35\%$ of published eccentric exoplanets cannot be distinguished from these two circular planets in 2:1 resonant configurations. This is a mathematical degeneracy that complicates the analysis of radial velocity observations and can only be overcome with high-precision, high-cadence data.

\subsection{Measuring precise radial velocities}

\begin{figure}
	\centering
	\includegraphics[width=\textwidth]{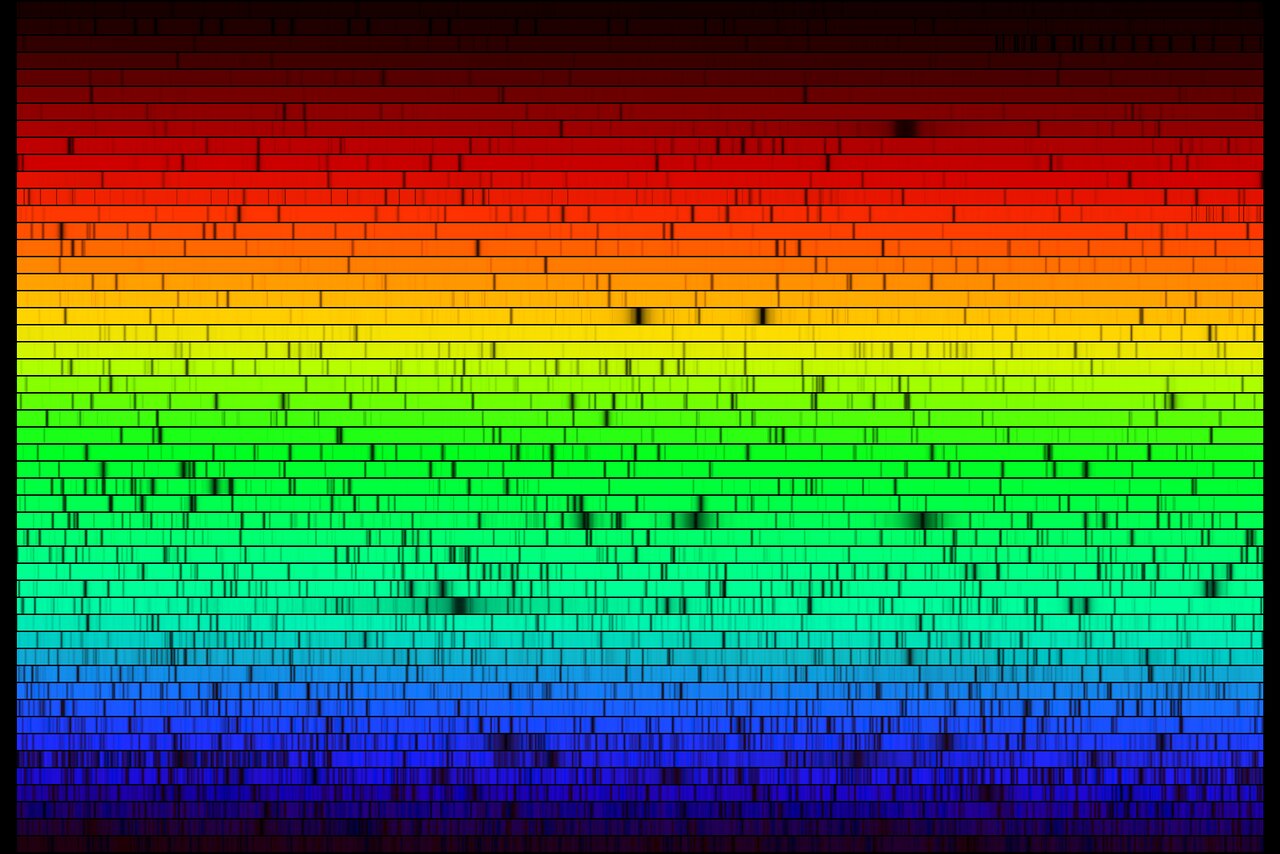}
	\caption{The solar spectrum. The dark vertical lines in the spectrum are absorption lines. Credit: \href{https://noirlab.edu/public/images/noao-sun/}{NOAO/NSO/Kitt Peak FTS/AURA/NSF}}
	\label{fig:sun_spectrum}
\end{figure}

Atoms and molecules absorb light at specific wavelengths, which can be measured in a stationary reference frame called the ``rest frame''. In stellar atmospheres, these atoms and molecular species absorb the star's light at these specific wavelengths \citep{Gray1992oasp.book.....G}. When starlight is passed through a spectrograph, it is dispersed to produce the star's spectrum, which reveals dark features known as ``absorption lines'' (see Fig.~\ref{fig:sun_spectrum}). Due to the Doppler effect, if the star is periodically moving relative to the observer, the wavelengths of these absorption lines are shifted from their rest frame values. The magnitude of this shift can be determined using Eq.~\ref{eq:doppler_effect}.

Thanks to the development of high-resolution spectrographs coupled with stable wavelength calibration methods saved on sensitive Charge-Coupled Device (CCD) detectors, we have been able to use this technique for exoplanet research in the past 40 years. However, the Doppler shift in wavelength of a given stellar line due to an orbiting planet is much smaller than the pixel resolution of the spectrograph itself. For a typical cross-dispersed echelle spectrograph with a resolving power of $R\equiv\lambda/\Delta\lambda \sim 100\,000$, the wavelength shift of a single pixel corresponds to radial velocity change of 1\,km\,s$^{-1}$. To detect giant or Earth-mass planets one must measure shifts hundred to tens of thousands better than that. Therefore, precise radial velocities at the m\,s$^{-1}$ level rely on averaging and combining the individual velocities of many lines. This is why stars with spectral types between F5 to M5\,V (masses between 0.3 and 1.2\,$M_\odot$) are the best suited for this technique. Stars with very high effective temperatures ($T_{\rm eff} > 10\,000\,\mathrm{K}$) have spectra that show barely any spectral features since all chemical elements in their photospheres are ionized. Besides, such stars are typically fast rotators, which smears out spectral lines even more via rotational broadening. For very cool dwarfs ($T_{\rm eff} < 3200\,\mathrm{K}$), the presence of complex molecular bands \citep[mainly TiO, VO and metal hydrides;][]{1991ApJS...77..417K,2012RSPTA.370.2765A} makes the many available spectra lines packed, less contrasted, and overlapping; together with the fact that most of their bolometric flux falls into the near-infrared, putting demanding constraints on instrumentation (telluric removal, cryogenic components, etc.).

Wavelength calibration is perhaps the most important step towards measuring precise radial velocities. This is required to assign each pixel in the detector with a unique wavelength and create a much finer grid where sub-pixel Doppler shifts can be measured. There are two main strategies for accurate wavelength calibration in the radial velocity technique. 

The first one is the iodine method \citep[I$_2$C;][]{MarcyButler1992PASP..104..270M}. It is based in the observation of a reference spectrum --- a gas cell filled with iodine placed in the telescope beam before it enters the spectrograph --- that is superimposed to the stellar spectrum. In this way, the iodine lines are used as a stable wavelength reference. If both the iodine and stellar lines shift in the same way, the reason for the apparent Doppler effect is instrumental; on the other hand, if the stellar lines shift relative to the iodine lines, the change is astrophysical and inherent to the star. The iodine lines track all instrumental changes, which can be calibrated and corrected, making spectrographs using this technique relatively simple and cheap. However, the wavelength range to derive precise radial velocities is very limited because most of the spectral information from the iodine's is contained in the region between 500 and 620\,nm. Furthermore, light losses in the cell are of the order of 20--30\% making these spectrographs relatively inefficient regardless of their science goals.

The second strategy is to build a spectrograph where the light from a calibration unit is fed with an optical fiber to the instrument, also known as the  simultaneous reference technique \citep{Baranne1996}. The spectrograph is then enclosed in a pressure- and temperature-stabilized tank. The stellar spectrum and calibration reference are ingested in the spectrograph with fibers that follow a very similar optical path within the instrument. In this way, instrumental changes due to pressure and temperature variations inside the tank are monitored with the spectrum of the calibration reference. There are several calibration unit sources, being the most common hollow cathode lamps (thorium-argon being the most used, but also uranium-neon or tungsten; \citealt{Kerber2008}), a Fabry-P\'erot (FP) interferometer \citep{Wildi2010,SchaeferReiners2012,Terrien2021,Seifahrt2022}, or a laser frequency comb (LFC, \citealt{Steinmetz2008,LoCurto2012,Metcalf2019}). The gain in efficiency and precision is compensated by the complexity of the optical design and environment stabilization systems. 

\begin{figure}
	\centering
	\includegraphics[width=\textwidth]{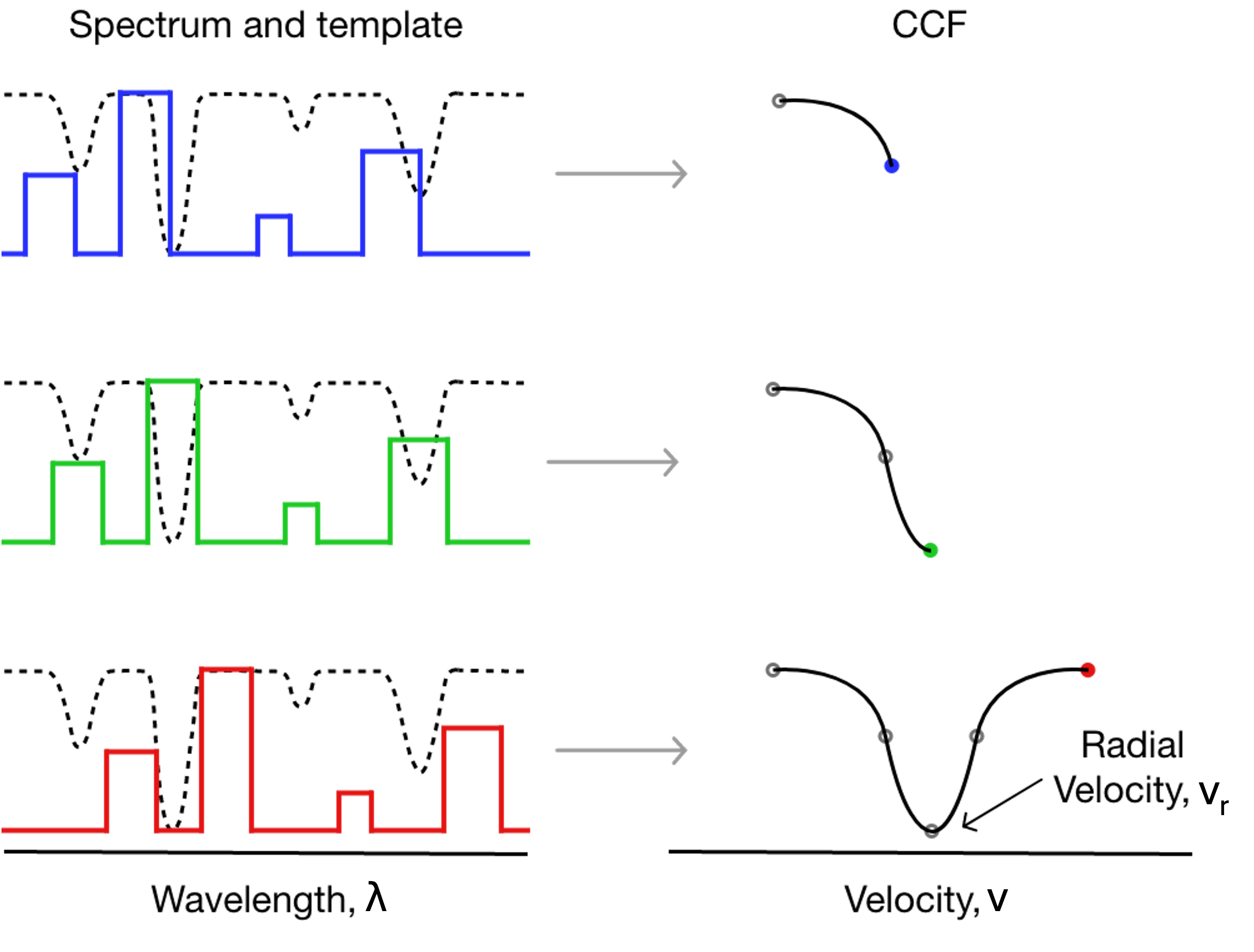}
	\caption{Illustration of the cross-correlation method to compute radial velocities. Left: Black dashed observed stellar absorption spectra and colored template spectra in wavelength space. Colors represent the continuous shift of the template across the spectrum from shorter (blue) wavelengths to longer (red) wavelengths as the black observed spectrum is fixed in place. Green represents the wavelength at which the template matches the observed spectra. Right: Resulting cross-correlation function (CCF) calculated at each step. The instantaneous radial velocity is then obtained from the center of a Gaussian fit to the CCF. Adapted from \citet{Roy2016}. }
	\label{fig:ccf_creation}
\end{figure}

Lastly, there are two main approaches to derive the final radial velocity value from all the individual stellar lines in the spectrum: the Cross-Correlation Function method \citep[CCF;][]{Baranne1996} or the template-matching technique \citep{TERRA,SERVAL}. The CCF method requires a weighted template spectrum (also known as ``masks'') that closely matches the star's spectral absorption lines. The template is then shifted through velocity space in steps across the observed spectra. The correlation between observed spectra and the template is calculated at each step (see Fig.~\ref{fig:ccf_creation}). This yields the cross correlation function, an inverted Gaussian shaped curve which represents the average shape of the observed stellar absorption lines. The radial velocity measurement corresponds to the centroid of a Gaussian function fitted to the mean-line profile. On the other hand, the template-matching technique minimizes the difference between the observed stellar spectrum and a high Signal-to-Noise ratio (S/N) template obtained by combining all the observations of the same star. This method has been demonstrated to be more accurate and precise for cool stars than the CCF technique \citep{SERVAL}, whose masks are difficult to build due to the blending of numerous adjacent lines and complicated continuum determination. Other techniques to extract precise radial velocity values leverage the information contained in individual lines (line-by-line method, \citealt{Dumusque2018A&A...620A..47D,LBL}), least-squares deconvolution \citep{Lienhard2022}, or the use of non-parametric Gaussian Processes \citep{Rajpaul2020}. These new methods aim to separate the purely translational radial velocity component produced by a planet from instrumental and/or astrophysical effects to reach higher accuracy and precision.

\subsection{Stellar activity and false positives}

\begin{figure}
	\centering
	\includegraphics[width=0.99\textwidth]{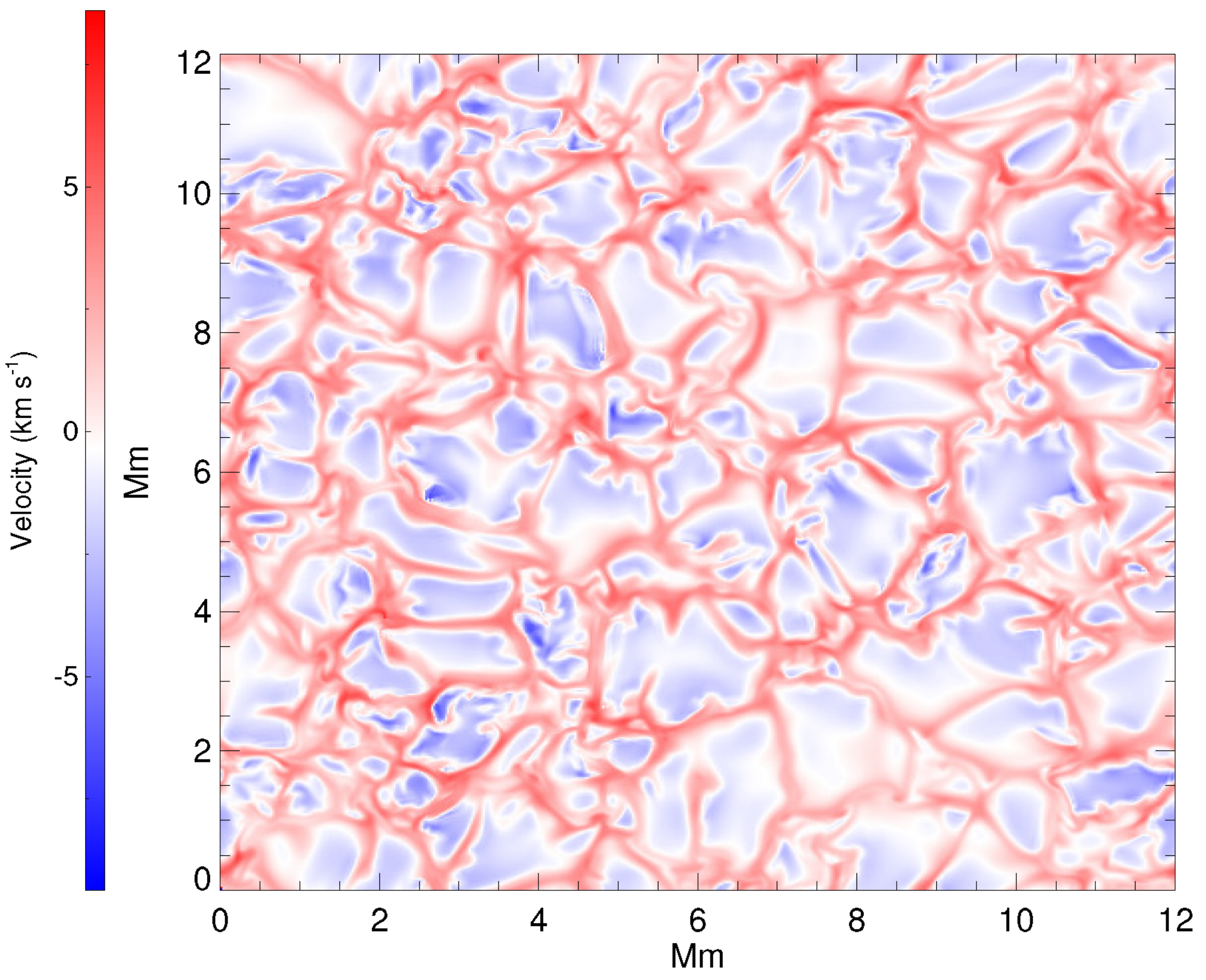}
	\caption{Net radial velocity effect for a 3D simulation of a Sun-like star at disc center. In the bright granules, plasma is moving upward, creating a blueshifted contribution to the radial velocity. On the other hand, in the dark intergranular lanes, plasma is moving downward, creating a redshifted contribution to the radial velocity. Since granules are more numerous than intergranular lanes, the net radial velocity of the spectral lines at the disc center is blueshifted, also known as convective blueshift net effect. Adapted from \citet{Cegla2019Geosc...9..114C}.}
	\label{fig:granulation}
\end{figure} 

\begin{figure}
	\centering
	\includegraphics[width=\textwidth]{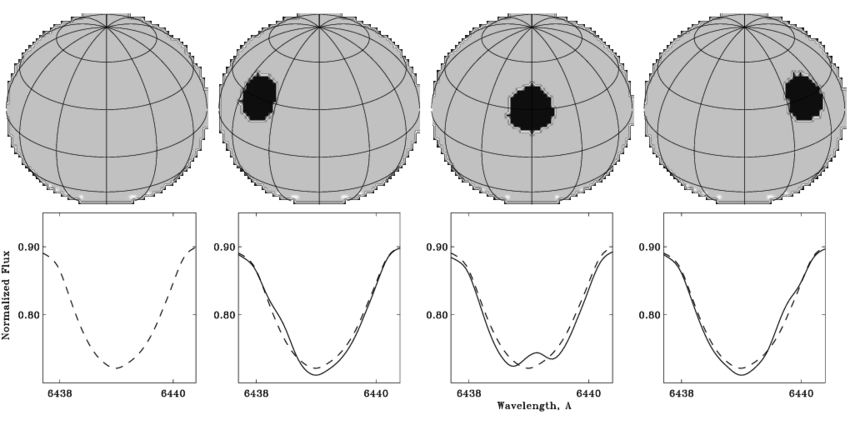}
	\includegraphics[width=\textwidth]{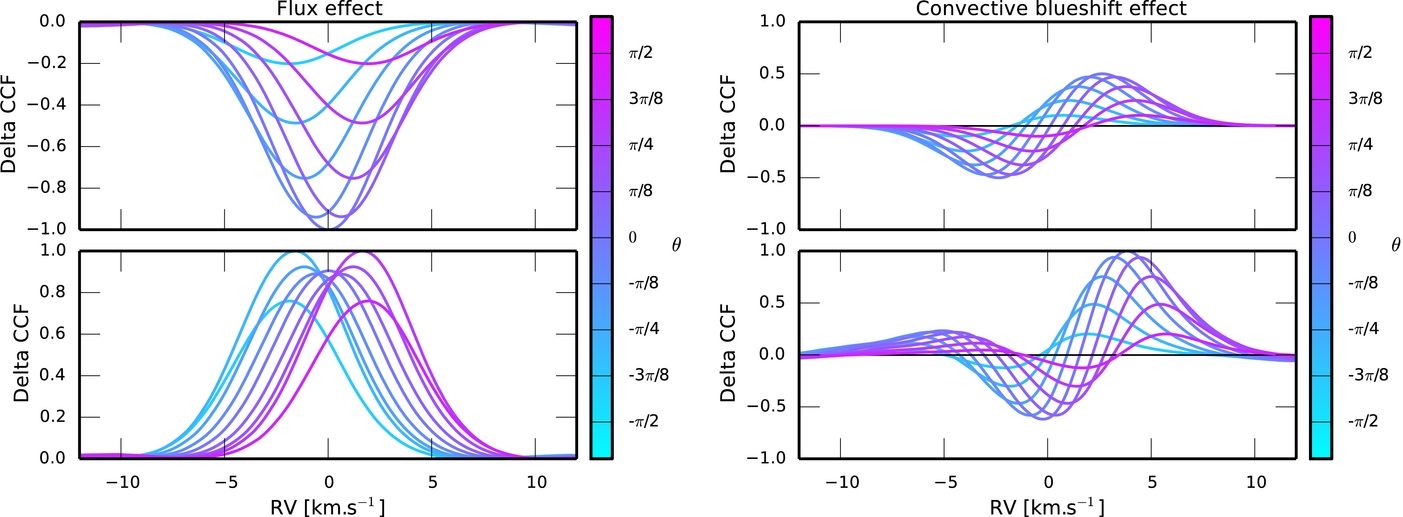}
	\caption{\textit{Top and middle}: Spectral line profiles for a model fast-rotating star with no spots (dashed line) and with a spot moving across the disk as the star rotates (solid line). \textit{Bottom}: Radial velocity effect induced by an equatorial spot or plage of size 1\% on a star seen equator-on as it rotates. The color indicates the phase of the stellar rotation. The left column shows the flux correction for a spot (top) or plage (bottom). The right column shows the convective blueshift correction when assuming Gaussian CCFs (top), and when assuming observed CCFs (bottom). The use of observed (non-symmetric) CCFs breaks the symmetry of the convective blueshift correction. Adapted from \citet{Berdyugina2005LRSP....2....8B} and \citet{Dumusque2014ApJ...796..132D}.}
	\label{fig:spot_RV}
\end{figure}

Since radial velocities are obtained from the Doppler shift of the star's spectral lines, any astrophysical process that affects the spectral line profiles can alter the radial velocity values regardless of the technique used. Stellar activity, and its influence on radial velocity measurements, is the main hurdle toward reaching the precision needed to detect ``Earth-Sun analogs'' and the most dynamic and rapidly-changing field of research using this technique. Here, we summarize the main mechanisms that impact radial velocity observations, some even mimicking the signatures of true planetary companions.

\subsubsection*{Stellar oscillations}
Pressure waves, also known as p-modes, propagate at the surface of Sun-like stars contracting and dilating the stellar surface over timescales of a few minutes. These oscillations create observable pulses in brightness and in the radial velocity with amplitudes between 10 and 400\,cm\,s$^{-1}$, depending on the star type and evolutionary stage \citep{SchrijverZwaan2000ssma.book.....S}. This phenomenon is predominantly observed in Sun-like stars and post-main sequence stars. Since radial velocity observations typically have sparse cadence (one measurement every few nights), this correlated noise appears as stochastic white noise typically referred to as RV jitter.

\subsubsection*{Granulation}
Surface granulation is a phenomenon that can only be directly measured on the Sun, but is expected to be present in other stars. The different phenomena of granulation are due to the convective nature of solar-type stars and can be found all over the stellar surface, except in active regions where convection is suppressed \citep{Dravins1982ARA&A..20...61D,Gray1992oasp.book.....G}. Convection creates time-variable asymmetries in the stellar absorption lines due to a combination of upward flowing, bright, hot, blue-shifted bubbles of plasma, known as \emph{granules}, that eventually cool, darken, and fall back down, red-shifting, into the surrounding regions, known as \emph{intergranular lanes} (Fig.~\ref{fig:granulation}, left). Since the granules are brighter and cover more surface area, they do not completely cancel out the intergranular lane contribution, which acts to depress the red wing of the combined line profile (Fig.~\ref{fig:granulation}, right). The larger contribution from the granules provokes an overall net blueshift for most absorption lines. The amplitudes of granulation phenomena are similar to the ones for p-modes \citep{Kjeldsen2005ApJ...635.1281K,KjeldsenBedding2011A&A...529L...8K}, being the reason that over the stellar disc much of the plasma upflows and downflows cancel out. However, the size of granules varies in size and lifetimes (from 8\,min to 1\,day), introducing RV variations which on the Sun are of the order of 0.5-1.0\,m\,s$^{-1}$ \citep{Meunier2015}. The effect of stellar oscillations and granulation in radial velocity observations can be typically averaged out with sufficiently long exposure times and targeted observing strategies \citep[e.g.,][]{Dumusque2011A&A...525A.140D,Chaplin2019AJ....157..163C,Luhn2023,Palumbo2024AJ....168...46P}, but it currently represents the biggest hurdle toward detecting RV signals of a few dozens cm/s in Sun-like stars \citep{Meunier2023A&A...676A..82M,Lakeland2024MNRAS.527.7681L}. 

\subsubsection*{Starspots, faculae and plages}
Dark spots, hot faculae, or bright plages are usually referred to as \emph{active regions} in the photospheres of stars. They exist in all the stars with outer convective zones and they are the main source of the stellar activity in spectroscopic observations. Starspots appear as dark spots on the surface of the stars. Local magnetic fields on the surface create them, suppressing the overturning convective motion and thus blocking or redirecting the flow of energy from the stellar interior outwards to the surface, appearing locally cooler (500--2000\,K) and darker. Faculae are bright spots that form in the intergranular lanes by the concentration of magnetic field lines. Plages are bright regions near starspots also associated with concentrations of magnetic fields. They are the counterpart of faculae in the chromosphere and are best seen in H$\alpha$.
Active regions cause photometric variations, radial velocity changes, and spectral line distortions due to their temperatures and the star's rotation. Groups of starspots form, move across the star's surface, and decay through turbulent diffusion. The stellar light blocked by dark spots includes Doppler-shifted continuum and line absorption features, influenced by rotation speed, axis inclination, and the spot's position. This creates a migrating bright bump in the absorption profile, shifting from blue to red as the spot co-rotates with the star, causing radial velocity variations and a convective blueshift effect (Fig.~\ref{fig:spot_RV}). These modulations when stable can mimic radial velocity variations induced by a true planetary companion, leading to false positives \citep[e.g.,][]{Queloz2001A&A...379..279Q,Huelamo2008A&A...489L...9H}. The amplitude of the Doppler shifts ranges between a few cm\,s$^{-1}$ to hundreds of m\,s$^{-1}$ with timescales ranging from a few to hundred days, depending on the projected rotational velocity of the star, the spectrograph resolution and both the size and temperature of the spot \citep{SaarDonahue1997ApJ...485..319S,Desort2007A&A...473..983D}.

\subsubsection*{Magnetic cycles}
Magnetic cycles such as the Babcock-Leighton solar dynamo cycle \citep{SchrijverZwaan2000ssma.book.....S} have been observed in spectroscopic monitoring programs of other stars \citep{Vaughan1981ApJ...250..276V,Makarov2010ApJ...715..500M}. The solar cycle in particular gives rise to a quasi-periodic 11-year cycle characterized by an increasing and decreasing number and size of active regions, coronal mass ejections, strong prominences, and levels of solar radiation \citep{Watson2011A&A...533A..14W}. Several studies have demonstrated that the majority of Sun-like stars exhibit magnetic activity cycles with periods ranging from 7 to 30\,years \citep{Duncan1991ApJS...76..383D,Baliunas1995ApJ...438..269B,Santos2010A&A...511A..54S}. The corresponding radial velocity signal from magnetic cycles can reach tens of m\,s$^{-1}$, thus influencing the ability to detect planetary companions using the radial velocity technique depending on the phase of the host's magnetic cycle.

Numerous approaches have been proposed to model and subtract the contribution of stellar activity from radial velocity measurements, this being an ongoing challenge that typically involves dense sampling and state-of-the-art modeling techniques. This is a vast topic that cannot be covered in this work. For more details about current strategies and challenges, we recommend the reader consult the reviews by \citep{EPRV-WG}, \citealt{Ford2024}, and Chapters 9--11 of \citet{Hatzes2019}.

\subsection{Difficulties in the detection of circumbinary planets}\label{sec:cbp_problems}

In addition to exoplanets orbiting single stars, exoplanets can also be found in and around binary star systems. It has been shown that systems with multiple stars exist as frequently those with single stars \citep{Tokovinin2014}, and to date, 375 binary stars and 45 hierarchical triple star systems have been found to host exoplanets \citep{Michel2024}. The majority of these planets orbit a single star in the system and the additional stars exist at a wide separation from the planet-hosting star. Although, perhaps the most iconic among the planet-hosting binary systems are those with circumbinary planets. These planets orbit around \textit{both} stars of a tight central binary system (with typical separations between the stellar pair $<10$~AU), exactly like the Star Wars planet Tatooine \citep{Lucas1977}. Circumbinary planets have long been thought to exist in nature \citep{Borucki1984, Schneider1994}, yet it was not until 2011 that the first confirmed circumbinary planet orbiting a main sequence binary was discovered. This was the discovery of Kepler-16~b using the transit method by \citet{Doyle2011} with data from the Kepler space telescope \cite{Borucki2010}. Since then, 15 main sequence circumbinary planets have been discovered using the transit and/or RV techniques.

However, despite the radial velocity method being extremely well established in the exoplanet field, only recently has it become possible to detect circumbinary planets using radial velocity observations. 
The first instance of this was HD~202206~b which was initially identified as a candidate circumbinary planet \citep{Correia2005}, though it was later reclassified as a brown dwarf \citep{Benedict2017} following a more precise determination of the systems inclination.
From the 15 known main sequence circumbinary planets, only 3 have been detected with radial velocities: TOI-1338/BEBOP-1~c \citep{Standing2023}, TIC~172900988~b \citep{Sairam2024a}, and the confirmation of Kepler-16~b \citep{Triaud2022}. 
However, this is not for lack of trying. Since the majority of these objects have been discovered by the transit method, the planet's mass (one of the planet's most fundamental parameters) could be estimated by measuring the eclipse timing variations in the lightcurves of the central binary. Unfortunately, this led to accurate mass measurements for only 6 of these transiting planets \citep{Standing2022}, calling for complementary observations such as radial velocities. 

To this end, a radial velocity survey known as The Attempt to Observe Outer-planets In Non-single-stellar Environments (TATOOINE) was launched in 2009 \citep{Konacki2009b, Konacki2010}. The survey observed 13 non-eclipsing double-lined spectroscopic binaries (SB2) with high-resolution spectrographs to discover circumbinary planets. SB2 systems are the most commonly found binaries where both stars are of similar brightness, meaning that the spectra of both stars are resolved in a single spectroscopic observation. This leads to complications in the radial velocity extraction process since the spectra from the stars themselves have vastly different radial velocities that blend and can produce blended CCFs with two inverted peaks. These blended CCFs can lead to a large residual scatter on the derived RV values \citep{Sairam2024b}. The TATOOINE survey reached a median residual Root Mean Squared (RMS) scatter of $\sim15\,m\,s^{-1}$ in their observations, down to $7\,m\,s^{-1}$ on their best target, HD~9939 \citep{Konacki2009b}. This low level of RMS scatter on SB2 binaries was an impressive feat at the time, proving no issue in ascertaining the binary orbital parameters. However, this level of RMS scatter can easily hide planetary signals in the data.

TATOOINE was unable to detect circumbinary planets in their SB2 sample and suggested that single-lined spectroscopic binaries (SB1) --- binaries where the secondary star is too faint for its spectra to be resolved in a single exposure --- should be targeted for future RV searches hoping to detect these elusive planets. The Binaries Escorted By Orbiting Planets (BEBOP) survey \citep{Martin2019} did just this, demonstrating sensitivity up to sub-Saturn mass planets \citep{Standing2022}. BEBOP also confirmed Kepler-16\,b independently measuring its mass \citep{Triaud2022}, and discovered the first circumbinary planet with radial velocity observations alone TOI-1338/BEBOP-1c \citep{Standing2023}. Following these successes, a new radial-velocity extraction process named \texttt{DOLBY} was developed. \texttt{DOLBY} uses Gaussian processes (GPs) to model time-varying components either in the SB2 spectra through spectral decomposition, or directly on the SB2 CCFs by modeling the CCF as a sum of a Gaussian function and a GP \citep{Sairam2024b}. This method significantly reduces the level of scatter produced when calculating the radial velocities of SB2 targets \citep{Sairam2024b}. \texttt{DOLBY} was then used to detect planet TIC-172900988\,b (initially identified by TESS in \citealt{Kostov2021}) independently using radial velocity observations \citep{Sairam2024a}, although their findings place the planet at a different orbital period from the transit solution. This difference could be due to the limited number of transits available from TESS at the time of discovery. The BEBOP survey is still ongoing with more candidate circumbinary planets awaiting confirmation with further RV observations \citep{Baycroft2024}.

Thanks to the dramatic improvement in the residual RV scatter of SB1 and SB2 data down to a few m\,s$^{-1}$ allowed by the \texttt{DOLBY} method, additional effects that can perturb a binary's Keplerian signal must be accounted for. These effects can be referred to as post-Keplerian effects, i.e., effects that can affect measured radial velocities in a non-Keplerian way. 
The short-term post-Keplerian signals which can affect the RV measurements of close binaries are gravitational redshift, transverse Doppler, light time effects, and tidal signals \citep{Kopal1980a, Kopal1980b, Kopeikin1999, Zucker2007, Konacki2010, Arras2012, Sybilski2013}. Another long-term post-Keplerian effect is that of the orbital precession of the binary orbit \citep{Baycroft2023}. These effects are important, and accounting for these effects can improve the sensitivity of the dataset and even reveal hidden low-amplitude planetary signals \citep{Baycroft2023}. In Section~\ref{sec:cbp_analysis}, we demonstrate how to account for these non-Keplerian signals in an example of the RV analysis of a close binary. For further information on the above effects, including their mathematical description, we refer the reader to the references above.

\section{Demographics of the detected planets}\label{sec:demographics}

\begin{figure}
	\centering
	\includegraphics[width=\textwidth]{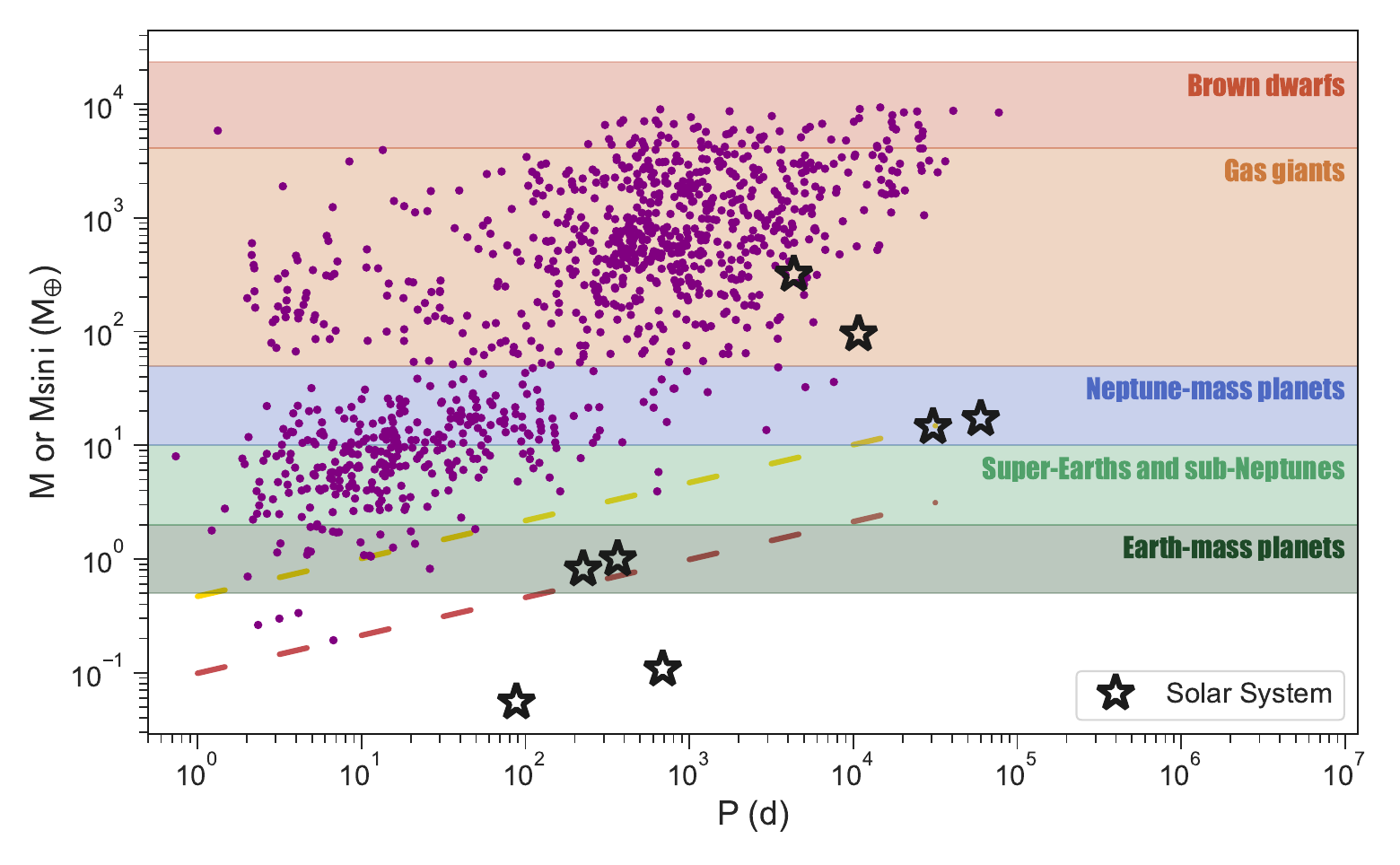}
	\caption{Mass-orbital period diagram of known exoplanets discovered by the radial velocity method. The planets of the Solar System are shown as black stars. The dashed line indicates the minimum mass detectable with a state-of-the-art radial velocity instrument (30\,cm/s internal precision) for a planet orbiting a Sun-like star (yellow) or a mid-type M dwarf (red). Data obtained from the NASA Exoplanet Archive in April 2025.}
	\label{fig:allrv_mass-period}
\end{figure}

\begin{figure}
	\centering
	\includegraphics[width=\textwidth]{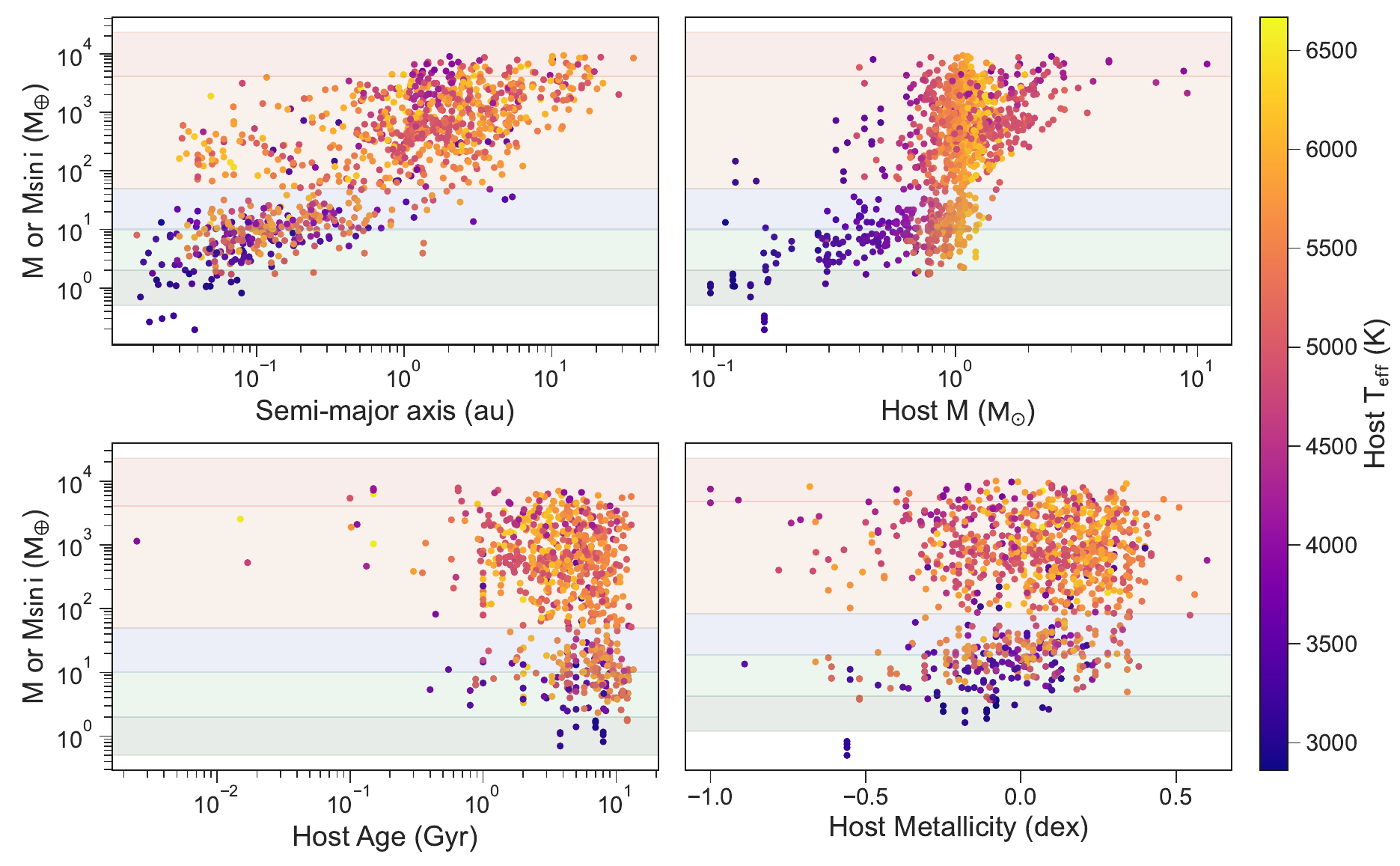}
	\caption{Mass of exoplanets discovered by the radial velocity method as a function of different system parameters. \textit{Top left}: orbital semi-major axis. \textit{Top right}: host stellar mass. \textit{Bottom left}: host age. \textit{Bottom right}: host stellar metallicity. In all panels, the data points are color coded by the host effective temperature. Data obtained from the NASA Exoplanet Archive in April 2025. The stellar parameters of the sample have not been derived homogeneously.}
	\label{fig:allrv_properties}
\end{figure}

\begin{figure}
	\centering
	\includegraphics[width=\textwidth]{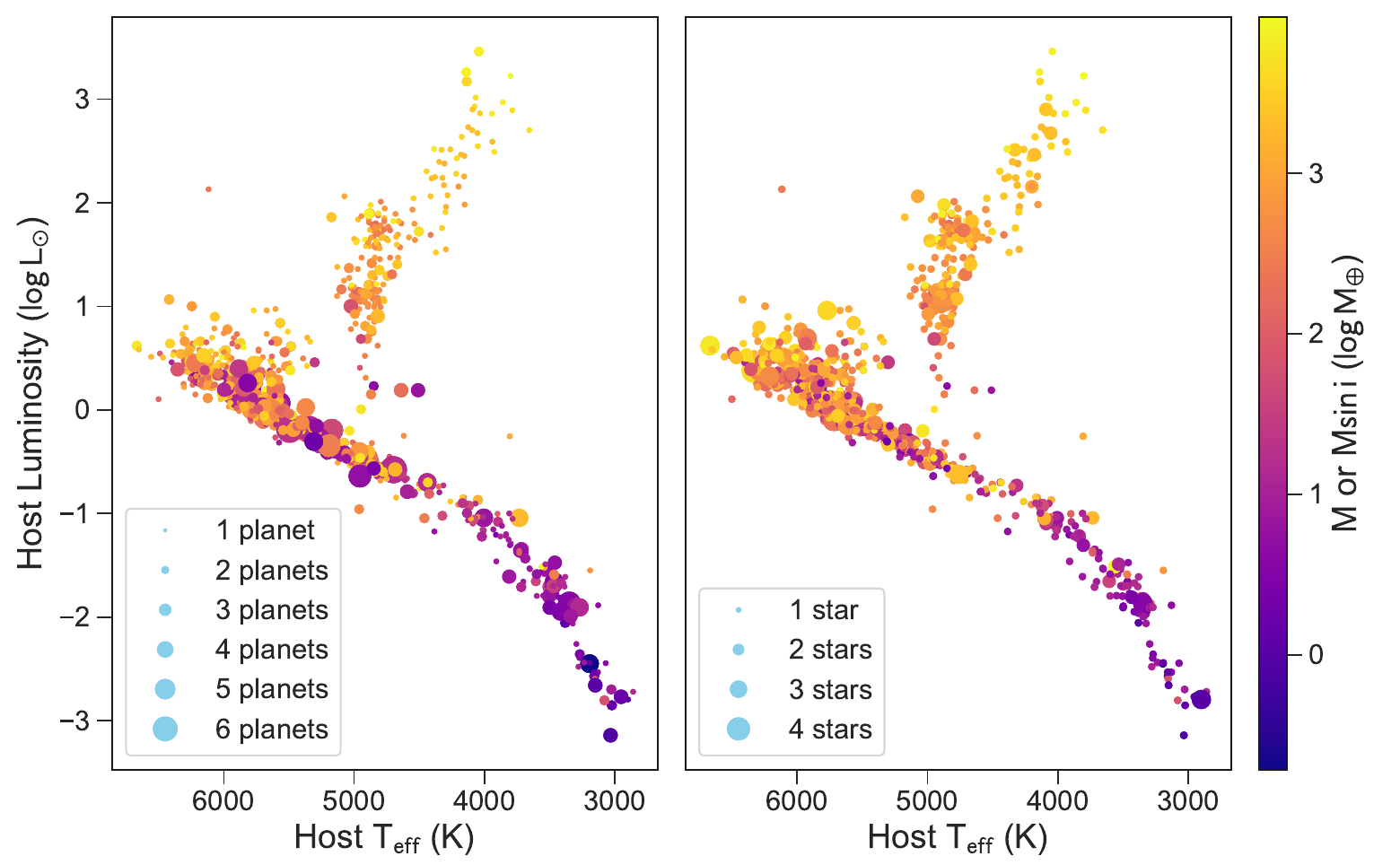}
	\caption{Hertzsprung-Russell diagram with all stars that have exoplanets discovered by the radial velocity technique. \textit{Left}: data point size indicates the number of planets in the system. \textit{Right}: data point size indicates the number of stars in the system. In both panels, the data points are color coded by the planet's mass. Data obtained from the NASA Exoplanet Archive in April 2025. The stellar parameters of the sample have not been derived homogeneously.}
	\label{fig:hr_mult}
\end{figure}

Over the past 30 years, astronomers have uncovered population-level trends in the properties of exoplanets. These trends relate to factors such as planetary mass, orbital period, and host star characteristics. This area of research, known as exoplanet demographics, aims to connect the measured properties of the observed exoplanet population to the physical processes driving planet formation and evolution. In this section, we will summarize some of the most relevant trends from radial velocity discoveries. For further details, including results from other techniques, we recommend the reader to consult, e.g., \citet{DemographicsASES3} and \citet{Cloutier2024}.

Figure~\ref{fig:allrv_mass-period} shows all planets discovered with the radial velocity technique in a mass-period diagram. There are multiple conclusions that can be extracted from this plot. First, with the exception of Jupiter, the exoplanet population does not overlap with the planets of the Solar System. Although we have detected some planets with masses below Earth and Venus and others at orbital distances longer than Uranus and Neptune, if we were to observe a Sun-like star with an exact copy of the Solar System planets, we would have been only able to detect Jupiter and Saturn with our current instrumentation. This mass-period diagram clearly exhibits the most important observational biases of the radial velocity technique. Following Eq.~\ref{eq:k}, the radial velocity semi-amplitude of a planet depends directly on the planet's mass and inversely on its orbital period and the host star's mass. Planets in shorter periods orbiting lower mass stars are easier to detect. The dashed lines in Fig.~\ref{fig:allrv_mass-period} indicate the mass of an hypothetical planet orbiting a Sun-like star (gold) or an M~dwarf (red) as a function of period if producing a constant radial velocity semi-amplitude of 30\,cm/s --- the state-of-the-art for current instrumentation. Besides this limitation of the technique, an additional bias can be identified in Fig.~\ref{fig:allrv_mass-period} where the abundance of planets with orbital periods beyond 10000~days appear to decline. This orbital period corresponds to roughly 30~years, the same time as the radial velocity instruments have been operating. This highlights the importance of long-term monitoring of stars over decades to enable the discovery of planets in the outer parts of their respective systems. 

In Fig.~\ref{fig:allrv_mass-period}, we can identify several demographic features. First, planets with masses between those or Earth and Neptune (super-Earths and sub-Neptunes) are very common at short-period orbits. \citet{Howard2010Sci...330..653H} and \citet{Mayor2011} measured the occurrence rates of these planets to be 10--30\% within 50-day orbits from surveys using HIRES \citep{HIRES} and HARPS \citep{HARPS}, respectively. Occurrence rates represent the average number of planets per star for a given range planetary and/or stellar properties. Most importantly, they account for the detection biases of a given survey, providing an estimate of the underlying, intrinsic exoplanet population. Second, gas giants are numerous, especially at orbital periods between those of Earth and Jupiter in the Solar System. \citet{Cumming2008} and \citet{Mayor2011} found that 10-20\% of solar-type stars ($0.8-1.2~M_\odot$) have giant planets with orbital periods shorter than a few years. The orbital distribution of giant planets appears to increase monotonically for Sun-like stars up to 2-3\,AU, where it starts to decrease again \citep{Suzuki2016,Fernandes2019,Wittenmyer2020,Fulton2021}. This distance corresponds to the approximate location of the snow line in the Solar System \citep{Morbidelli2016}. Finally, despite their observational advantages, hot Jupiters (incredibly numerous in transit discoveries) and planets with masses between those of Neptune and Saturn are intrinsically rare. The lack of Neptune-mass planets in short orbital periods, also known as the ``Neptune desert'', was first identified from the radial velocity surveys mentioned above and then confirmed by Kepler's transit survey \citep{Mazeh2016}.

Constraining the occurrence rate of Earth-like planets in the habitable zone of their host stars \citep{Kasting1993,Kopparapu13} is one of the ultimate goals of demographic studies. Most studies able to constrain this value, typically referred to as ``eta Earth'' ($\eta_\oplus$), used transit observations from Kepler \citep[e.g.,][]{KunimotoMatthews2020,Bergsten2023}. Radial velocity surveys have estimated that $\eta_\oplus$ is approximately equal or smaller than 20\% for early-type M~dwarfs \citep{Bonfils2013,Pinamonti2022}. Improved occurrence rates for terrestrial planets orbiting M~dwarfs are expected from the ongoing CARMENES Legacy survey \citep{Ribas2023} and the blind searches from the MAROON-X \citep{Seifahrt2018} and NIRPS \citep{NIRPS} instruments. Pushing towards Sun-like stars to derive $\eta_\oplus$ from radial velocity requires dedicated surveys with extreme precision and long monitoring. {Chapter~10 describes some of those upcoming projects, including the ``100 Earths survey'' from EXPRES \citep{Brewer2020}, the ``Terra Hunting Experiment'' (THE) using HARPS3 \citep{Thompson2016}, and the ``Second Earth Initiative Spectrograph'' (2ES) for the 2.2m ESO/MPG telescope in Chile \citep{Stuermer2024}. }

\subsection{Host star demographics}

Figure~\ref{fig:allrv_properties} shows the population of exoplanets discovered by radial velocities as a function of different properties of the stellar host. One of the most noticeable trends is between stellar metallicity (using [Fe/H] as proxy of the total metallicity) and the presence of gas giants (bottom right panel in Fig.~\ref{fig:allrv_properties}). This trend was discovered as early as 1997, when \citet{Gonzalez1997} noticed the relatively high metallicity of the host stars of the four giant exoplanets known at the time. \citet{FischerValenti2005ApJ...622.1102F} confirmed, by measuring the spectroscopic parameters of more than 1000 solar-type stars in different planet search programs, that there is a smooth increase in the occurrence of giant planets as a function of stellar metallicity. For metal-rich stars ($\mathrm{[Fe/H]} >+0.3$), the authors found that 25\% of them had planets. The giant planet-metallicity relation has been measured in radial velocity surveys of stars of all spectral types (M dwarfs, solar-type stars, and evolved stars) and confirmed with transit surveys \citep{SchlaufmanLaughlin2011,Everett2013}. On the other hand, the occurrence rate of low-mass planets appears independent of stellar metallicity \citep{Sousa2008,Mayor2011}.

As a function of host mass (top right panel in Fig.~\ref{fig:allrv_properties}), we can identify that gas giants are more common for more massive stars, while low-mass stars most host Earths and super-Earths and rarely any gas giants. Surveys focusing on low-mass stars find that M~dwarfs host at least 10 times less giant planets and 2-3 times more low-mass planets (Earths, super-Earths, and sub-Neptunes) than Sun-like stars \citep{Endl2006ApJ...649..436E,Bonfils2013,Sabotta2021,Pinamonti2022}. For giant planets with semi-major axes shorter than 2.5\,AU, their occurrence rate increase with host star mass, peaking at approximately $1.7\,M_\odot$ \citep{Reffert2015A&A...574A.116R,Wolthoff2022}. Core accretion models \citep{Pollack1996} offer a unified picture to explain all these trends. The timescale for core growth is directly related with the amount of material available in the disk. As stellar metallicity is a tracer of the solids in the protoplanetary disk, metal-rich stars have readily available more material to form giant planets \citep{Ikoma2000,KokuboIda2002}. Furthermore, the stellar mass is directly proportional to the mass of the protoplanetary disk, explaining why more massive stars host giant planets more frequently \citep{Andrews2013,Lozovsky2021A&A...652A.110L}. The decline after $1.7\,M_\odot$ can be explained by planet engulfment \citep{Stephan2018} and the short life time of A stars compared to Sun-like stars \citep{Ribas2015}. But, as a function of host age, the demographics of radial velocity exoplanets are extremely incomplete. Young stars are extraordinarily active and fast rotating. The enhanced levels of stellar activity and rapid rotation lead to decreased sensitivity to planetary signals. Our understanding of young planet demographics is mostly restricted to direct imaging (Chapter~5) and transit (Chapter~3) techniques.

Finally, not every star is amenable for precise radial velocity measurements. Stars with very high effective temperatures ($T_{\rm eff} > 10 000$\,K) have optical and near-infrared spectra that barely show spectral features (beyond H- and He-lines) since all chemical elements in their photospheres are ionized. Besides, such stars are typically fast rotators, which smears out spectral lines even more via rotational broadening. For very cool dwarfs ($T_{\rm eff} < 3000$\,K), the presence of complex molecular bands make the many available spectra lines packed, less contrasted and overlapping; together with the fact that most of their bolometric flux falls into the near infrared, putting demanding constraints on instrumentation (telluric removal, cryogenic components, etc.). Figure~\ref{fig:hr_mult} shows a Hertzsprung-Russell diagram with all the stars hosting exoplanets discovered by the radial velocity technique. Stars with effective temperatures between 3000 and 6300 K (both main sequence and giant stars) are the best suited for radial velocity exoplanet searches. The diagram also shows the role of planet and stellar multiplicity on these discoveries. Despite the number of discoveries of single systems and orbiting single stars are overwhelmingly larger, it can be seen that multi-planetary systems are common among all main sequence stars and that many of the known planets orbiting giant stars are in multiple star systems. For planet multiplicity, \citet{Wright2009ApJ...693.1084W} showed that about 28\% stars with one radial velocity planet host additional companions while \citet{Mayor2011} reported that 40\% of systems with low-mass planets are multiple. In addition, multi-planetary systems tend to have lower eccentricities than single-planet systems, suggesting violent dynamical histories for the latter \citep{Limbach2015}. 

For planets in multiple stellar systems the stellar companions can inhibit planet formation in some regions of parameter space in the system.  This has been studied extensively \citep{Dvorak1989, Mardling1999}, and numerical simulations find unstable regions immediately surrounding close binaries, and external to planets orbiting one star of wider binaries \citep{Holman1999}. For close binaries, the unstable region immediately surrounding both stars extends to $\approx~3 \rm{a_{bin}}$ \citep{Martin2017a}, depending on the eccentricity of the binary stars \citep{Holman1999}. For wide binary systems, planetary orbits can also be altered via mechanisms such as Kozai-Lidov cycles \citep{Kozai1962, Lidov1962}. For example a planet orbiting a single star in a wide binary, if all objects begin on circular orbits and the outer star is on a sufficiently inclined orbit, then high amplitude oscillations in eccentricity and mutual inclinations are induced in the primary star-planet system \citep{Martin2016}. Although, \cite{Martin2016} found that these cycles do not occur in circumbinary planet configurations where the angular momentum from the planet is negligible \citep{Martin2018}. Therefore, when analyzing radial velocity data for circumprimary/circumsecondary (s-type) systems, one should expect a non-zero eccentricity and mutual inclination for the planetary orbit. These theoretical and empirically tested unstable regions allow us to place strict limits on the priors of the parameters of interest in our radial velocity fits (see Section~\ref{sec:cbp_analysis}).

\cite{Eggenberger2007, Eggenberger2011} searched for stellar companions with direct imaging (Chapter~5) in systems where radial velocity observations had previously identified a planetary companion. They found that planet-hosting stars are less likely to have stellar companions at distances closer than 100~AU, compared to field stars in their sample without any detected planet. However, there was no noticeable difference between the samples for stellar companions between 100-200~AU.

\section{Dataset analysis example}\label{sec:Analysis}

\subsection{Single star example (55 Cnc)}

\begin{figure}
	\centering
	\includegraphics[width=\textwidth]{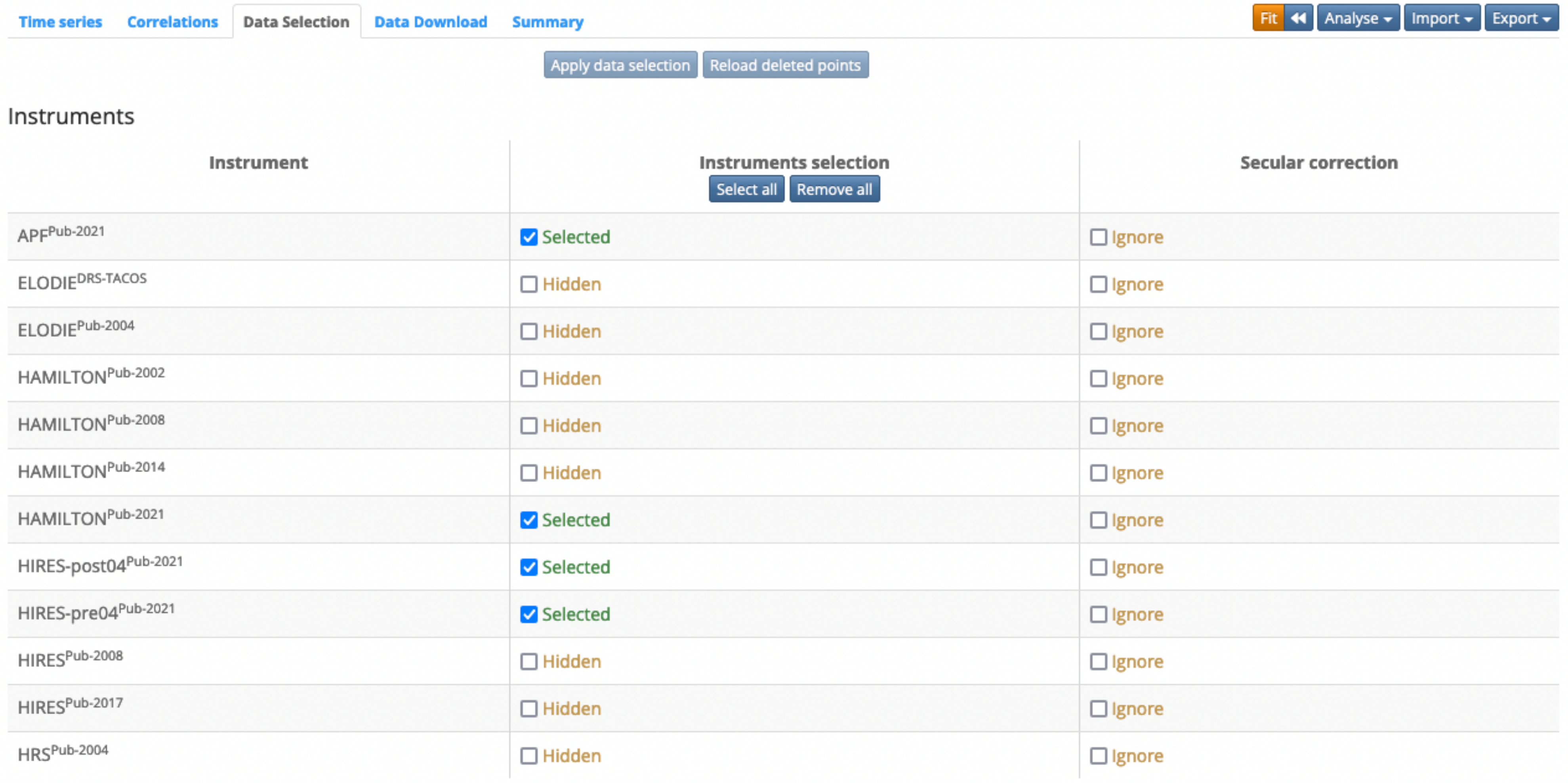}\\
	\includegraphics[width=\textwidth]{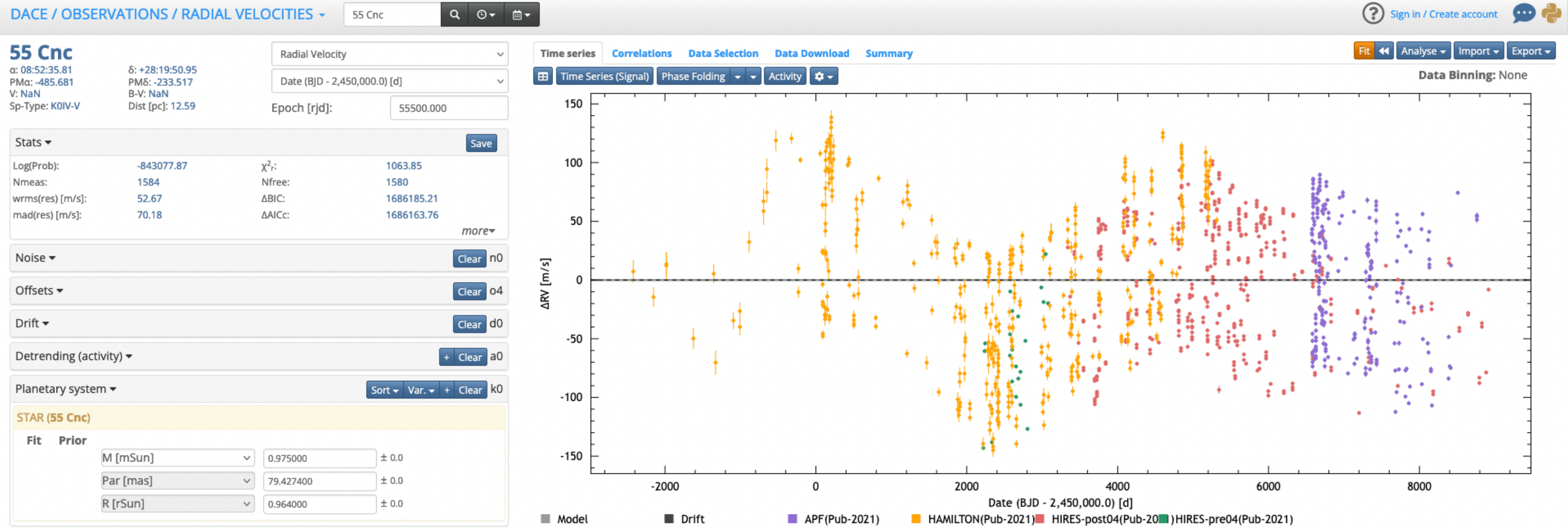}
	\caption{Top: Data Selection tab from DACE where we select the instruments from the analysis of 55 Cnc presented in \citet{Rosenthal2021}. Bottom: Overview of the 55 Cnc data, including stellar properties, basic statistics of the data set, and a plot showing the time series of the radial velocity data.}
	\label{fig:55Cnc_1}
\end{figure}
\begin{figure}
	\centering
	\includegraphics[width=\textwidth]{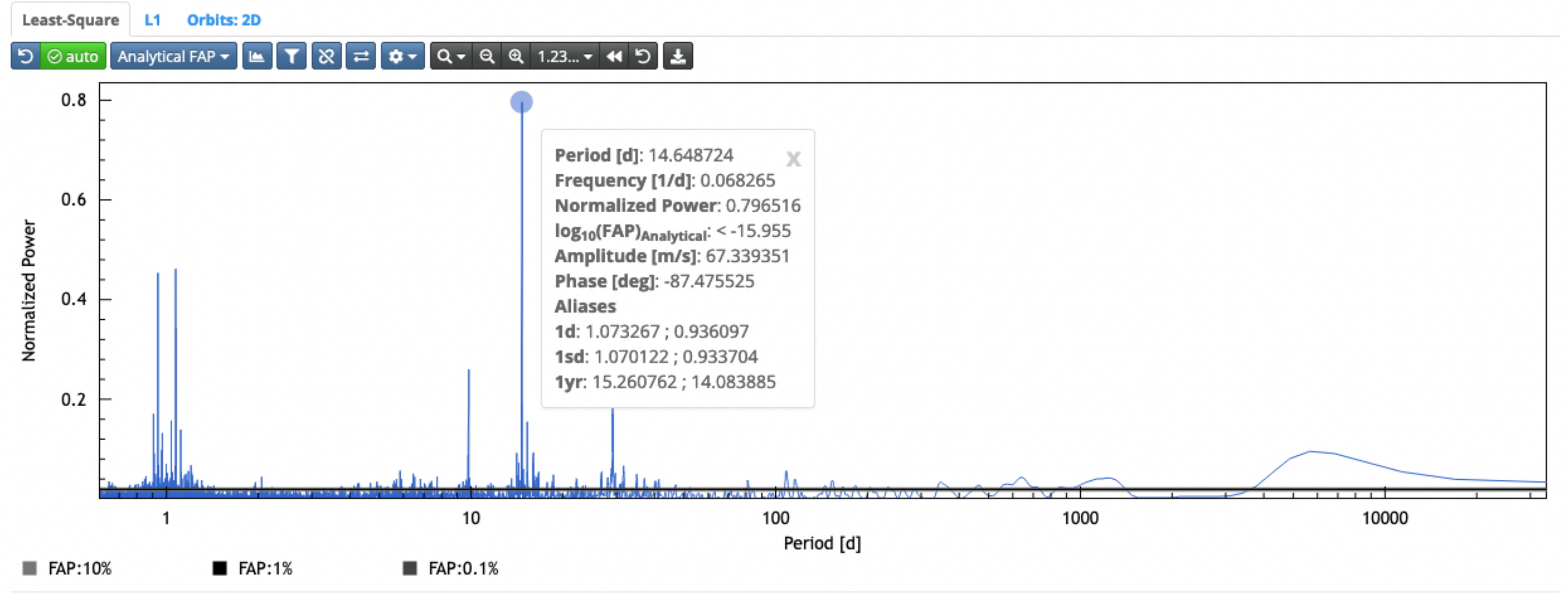}
	\caption{GLS periodogram of the 55 Cnc dataset. Hovering the cursor over the strongest peak in the periodogram gives the information shown in the white box, which includes its period, frequency, power, false alarm probability, and most common aliases.}
	\label{fig:55Cnc_2}
\end{figure}
\begin{figure}
	\centering
	\includegraphics[width=\textwidth]{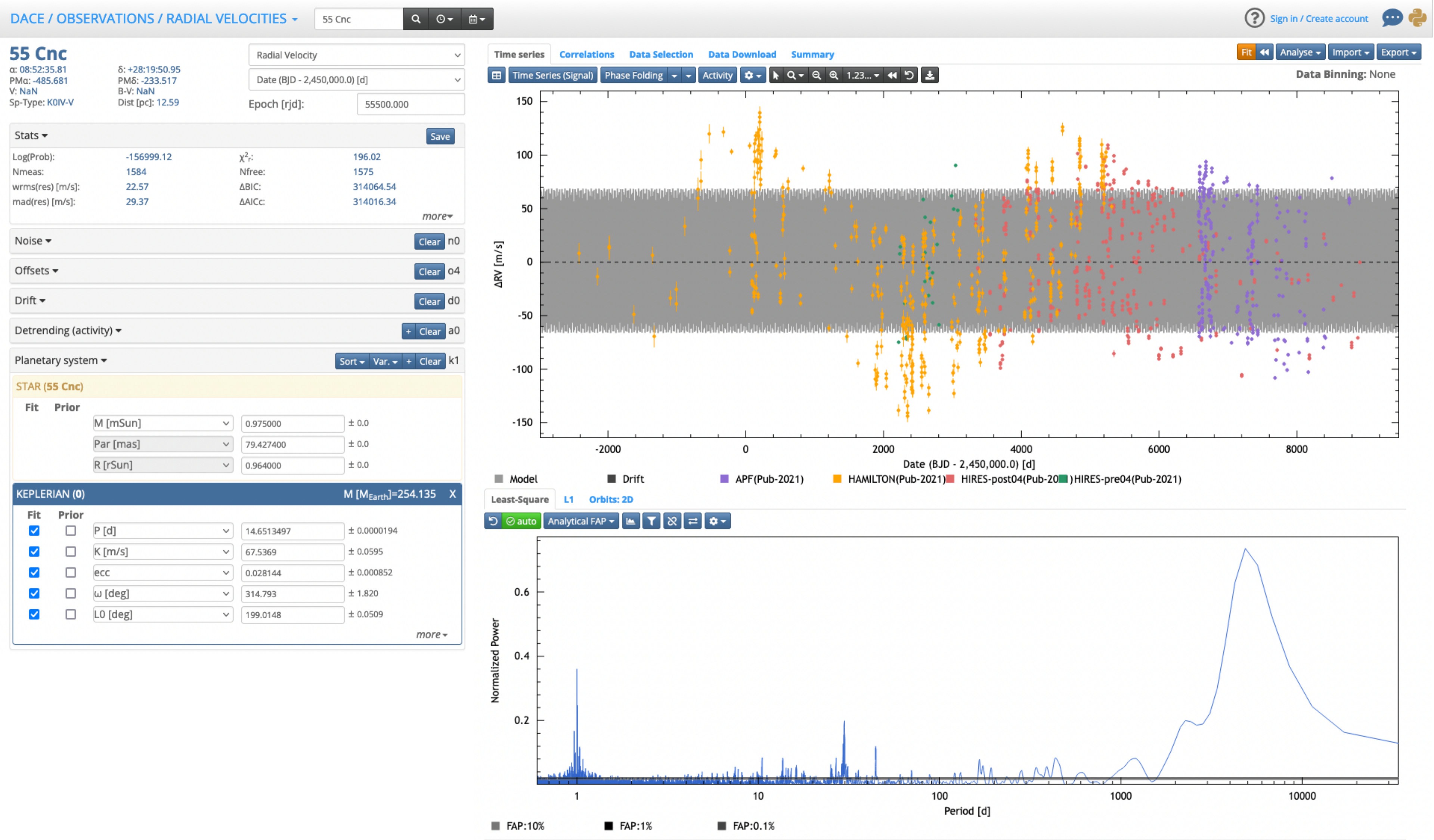}
	\caption{Screen from DACE after fitting the 14.65-day signal with a Keplerian orbit. The bottom left corner shows the best-fit parameters of the Keplerian model. The top right plot shows the time series with the best-fit model superimposed. The bottom right plot shows the GLS periodogram of the residuals from the fit.}
	\label{fig:55Cnc_3}
\end{figure}
\begin{figure}
	\centering
	\includegraphics[width=\textwidth]{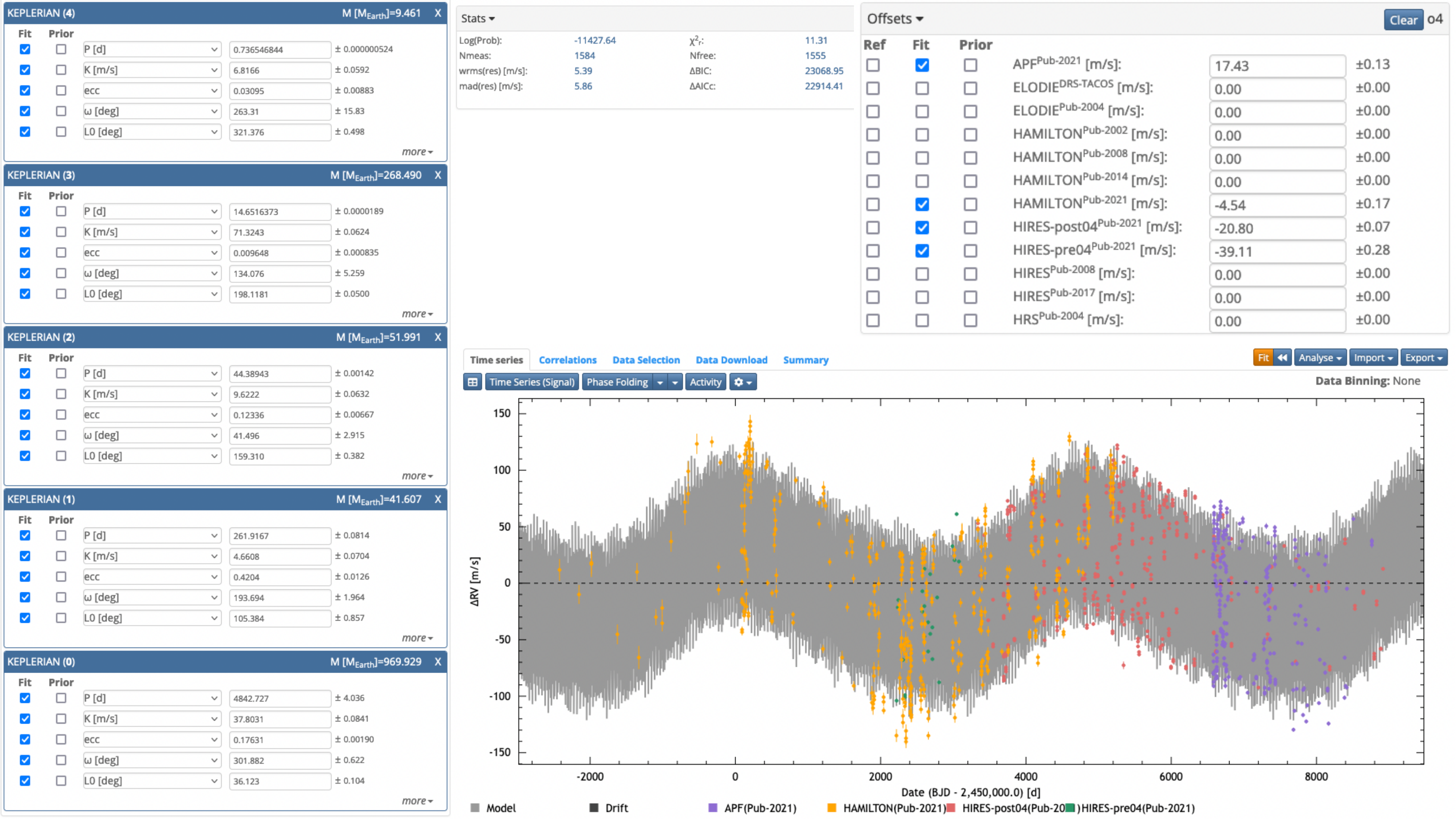}
	\caption{Final model for the 55 Cnc data, which includes five Keplerian orbits (whose parameters are shown on the left) and four instrumental offsets (best-fit values on the top right panel) for a total of 29 free parameters. The bottom right plot shows the data overlaid by the full best-fit model.}
	\label{fig:55Cnc_4}
\end{figure}

In this section, we analyze a subset of radial velocity observations of the iconic star 55~Cnc taken with the Hamilton, HIRES, and APF instruments between 1989 and 2020 as described in \citet{Rosenthal2021}. We make use of the web platform DACE \citep{DACE} at \url{https://dace.unige.ch/} to access the publicly-available radial velocity observations of this star, carry out a signal search using periodograms, and model the data with Keplerian orbits. The host star 55 Cnc is a K0 main sequence star at a distance of 12.6\,pc hosting a multi-planetary system that includes super-Earths, sub-Neptunes, and giant planets at both short- and long-period orbits \citep{Butler1997,Marcy2002,McArthur2004,Fischer2008}. Here, we describe a basic analysis of the data for educational purposes as more detailed analyses of this system have been published in the literature that the reader can consult \citep[e.g.,][]{Bourrier2018,Rosenthal2021}.

First, we use DACE to retrieve and select the observations from the star that we are interested in. In the \texttt{Target search} box in \url{https://dace.unige.ch/radialVelocities/}, we write ``55 Cnc'' and click the search icon. The app will load the observations and issue the following warning ``Big data set''. To speed up the computation time of our analysis, we restrict our dataset to those analyzed in \citet{Rosenthal2021}. To do that, we select the \texttt{Data Selection} tab and select only the instruments APF (Pub-2021), Hamilton (Pub-2021), HIRES-post04 (Pub-2021), and HIRES-pre04 (Pub-2021) as shown in the top panel of Fig.~\ref{fig:55Cnc_1}. This selection ensures the same data and data reduction version as in \citet{Rosenthal2021}. Then, we can select the \texttt{Time series} tab to see the data as a function of time and some basic properties of the dataset in the \texttt{Stats} box as shown in Fig.~\ref{fig:55Cnc_1}. The data shows a clear periodic modulation with a period of approximately 5000\,days and additional short-term variability with a peak-to-peak amplitude of 150\,m\,s$^{-1}$. 

Most researchers searching for exoplanets in RV data use periodograms as a tool to find period signals in unevenly spaced time series. DACE includes two types of periodograms: a Generalized Lomb Scargle periodogram \citep{GLS} and an $\ell$-1 periodogram which is less susceptible to aliasing \citep{Hara2017}. For this analysis, we will make use only of the GLS periodogram as it is computationally faster. Figure~\ref{fig:55Cnc_2} shows the GLS periodogram of the full dataset. The highest peak is at a period of 14.648\,days, with two other peaks near 1\,day with approximately half the normalized power. These peaks are the 1-day aliases of the 14.648\,day signal. Aliases are a natural consequence of the cadence of the observations and the presence of gaps in the time series. The neighboring peaks at 14.08 and 15.26\,days are due to 1-year aliases related to the visibility of the star during the year. A very detailed and informative text on periodograms, aliasing, and everything related to the analysis of RV data in frequency space can be found in Chapter 7 of \citet{Hatzes2019}.

We model the 14.648-day peak with a Keplerian orbit within DACE. We consider this peak to be statistically significant as its analytical false alarm probability in the GLS periodogram is much smaller than 0.1\% \citep{Baluev09}. Other criteria can be used, especially when using different periodograms, but this is a common threshold used in the field. The user can fit a Keplerian to this peak by clicking on and choosing the action ``Add Keplerian'', which will automatically add a Keplerian orbit in the box \texttt{Planetary system} on the left of the screen. The orbital parameters of the Keplerian are initialized with the period found on the GLS periodogram, then minimized using the L-BFGS-B algorithm included in \texttt{scipy.optimize.minimize} to find the best-fit parameters. Figure~\ref{fig:55Cnc_3} shows the best-fit parameters of the Keplerian model, which correspond to a planet with a mass of 254\,$M_\oplus$ (0.8\,$M_J$) in a nearly circular 14.65-day orbit. The top right plot shows the RV time series with the model, which lacks an additional long-term signal. This can be confirmed on the bottom right panel, where the GLS periodogram of the residuals of the fit (which is automatically generated every time a component is added to the model) shows a very strong peak with a period of approximately 5000 days. Despite the model being incomplete, there is a significant decrease in the weighted RMS of the residuals from the original data (52.7\,m\,s$^{-1}$) to the 1-Keplerian model (22.6\,m\,s$^{-1}$). This indicates that the 1-Keplerian is a much better model than the fiducial one only accounting for the different instrumental offsets. DACE also reports additional metrics such as the reduced $\chi^2$, Bayesian Information Criteria, and Akaike Information Criteria to perform basic model comparison and evaluate the significance of the modeled signals.

We can follow this process iteratively --- adding Keplerian orbits to each significant peak in the periodogram of the residuals --- until the change in one of those statistical metrics becomes statistically insignificant or there is no longer an improvement in the scatter of the residuals. For the sake of brevity, we show the complete model for 55 Cnc in Fig.~\ref{fig:55Cnc_4} after adding 4 more signals we find highly significant in the GLS of the residuals. We fit the data with five Keplerians plus additional free offsets for each instrument. The solution is very similar to the one found by \citet{Rosenthal2021}, which corresponds to a system with five planets at 0.736, 14.6, 44.4, 261, and $\sim$4800 days. The orbital architecture of the system is very peculiar as it has two very massive planets at short- and long-orbital periods, making one of the few known systems with a hot and a cold Jupiter. The innermost planet at 0.736\,days is the infamous 55~Cnc~e, a lava world that has been the subject of a multitude of programs trying to detect its atmosphere (if it has one) with disparity of results \citep{Demory2012,Demory2016ApJ...825L..25D,Tsiaras2016,Deibert2021,Hu2024,Patel2024}. 

Finally, it is important to note that it would be possible that additional signals are still present in the data or that the modeled signals do not correspond to bonafide planets, but rather instrumental or stellar variability. Deciding how many planets are present and what their orbital parameters are requires careful numerical methods, checking for appropriate convergence and statistical robustness. Model comparison and cross-validation are fundamental aspects to account for when analyzing radial velocity observations, especially in the presence of stellar variability, telluric contamination, and instrumental systematics. Another important aspect is a correct estimation of the uncertainties in the model parameters. To do that, methods such as Markov Chain Monte Carlo (MCMC) or nested sampling are typically used to obtain posterior distributions that account for correlations between parameters and sample the full parameter space of allowed solutions. A detailed assessment of all these aspects is at the forefront of exoplanet research and beyond the scope of this analysis. We recommend the reader to check the review by \citet{HaraFord2023AnRSA..10..623H} on this topic.

\subsection{Circumbinary planet example (TIC-172900988)}\label{sec:cbp_analysis}

\begin{figure}
	\centering
	\includegraphics[width=\textwidth]{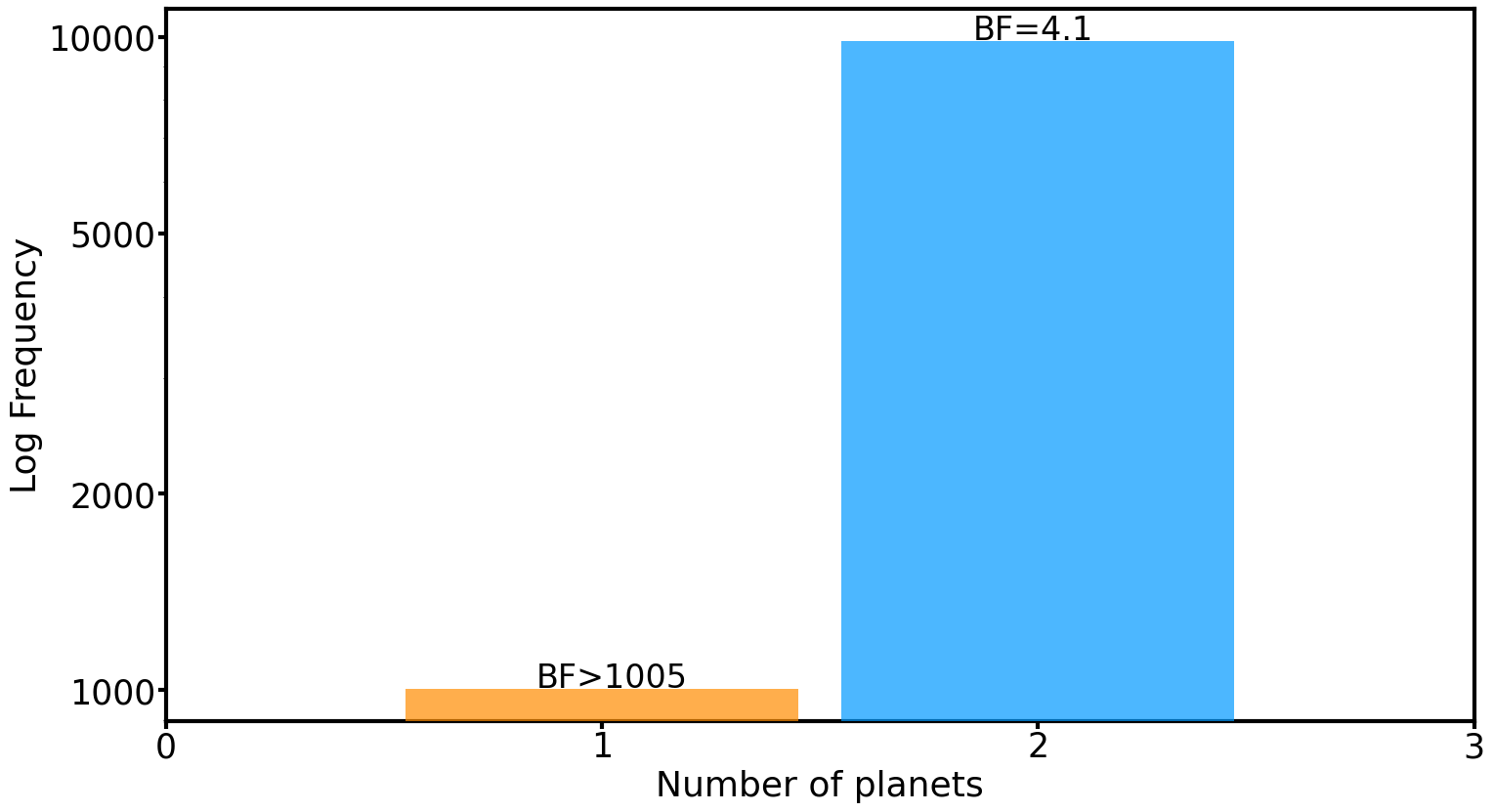}
	\caption{Histogram of the number of posterior samples with $\rm{N_{p}}$ planetary signals from a \texttt{kima} run on the TIC-1729 dataset. The chosen $\rm{N_{p}}$ here is 1 as it has the highest Bayes factor, denoted by the orange bin.}
	\label{fig:TIC-1729_Np_histogram}
\end{figure}

\begin{figure}
	\centering
	\includegraphics[width=\textwidth]{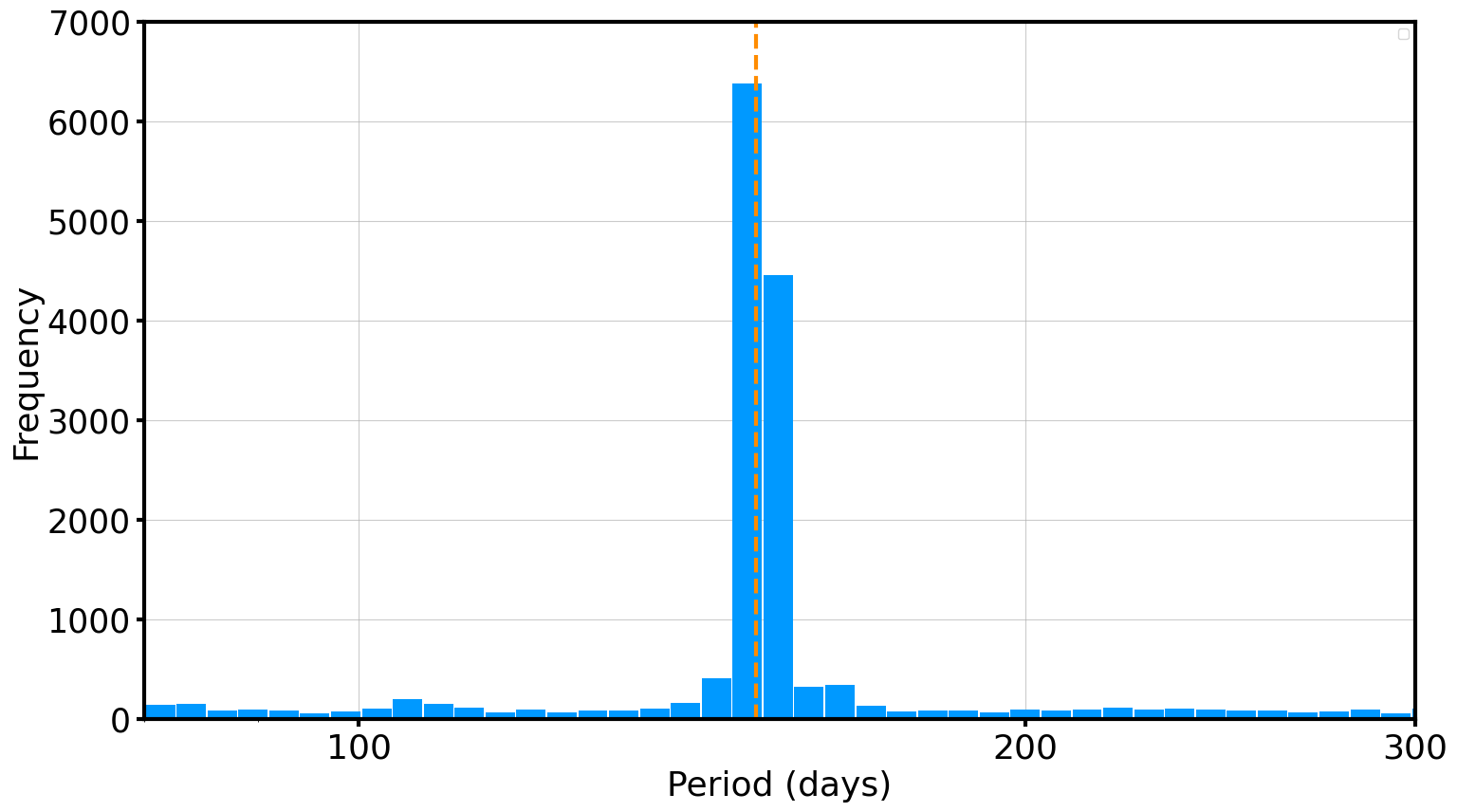}
	\caption{Posterior density function of the orbital period samples from a \texttt{kima} run on TIC-1729 data. The clear peak corresponds to the orbital period of the posterior sample with the highest likelihood. This period corresponds to the planet TIC-1729b (151.2\,d) \citep{Sairam2024b}, indicated by the vertical orange dotted line.}
	\label{fig:TIC-1729_periodogram}
\end{figure}

\begin{figure}
	\centering
	\includegraphics[width=\textwidth]{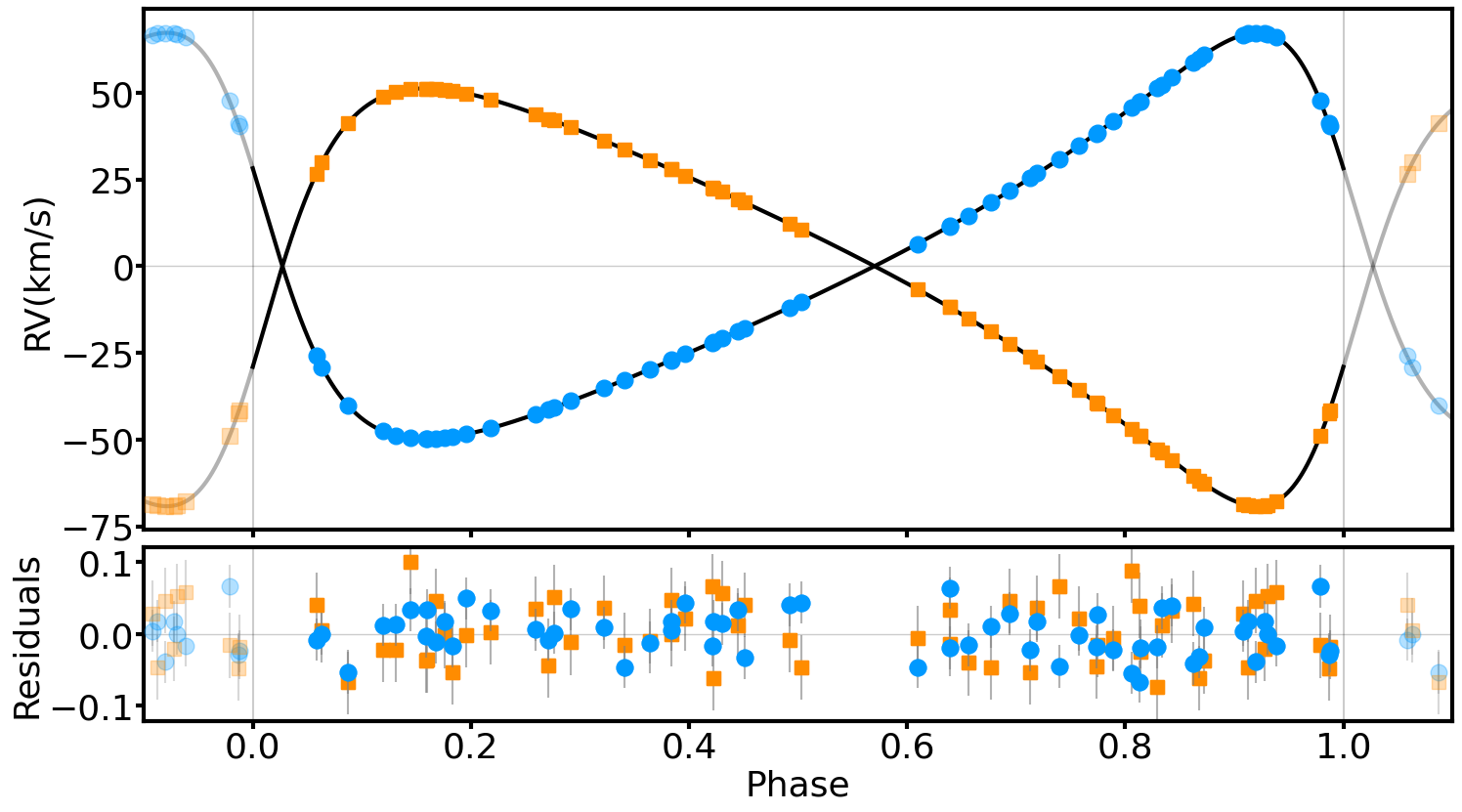}
	\caption{Phased radial velocity plot of highest likelihood \texttt{kima} sample showing the radial velocity motion of both binary stars in TIC-1729 along with their associated residuals. The black line is the highest likelihood Keplerian RV signal fit to the data. RV data from the primary star is represented by the blue circles and the secondary star B by the orange squares. RV data is from \citet{Sairam2024b} obtained using the SOPHIE spectrograph \citep{Perruchot2008}.}
	\label{fig:TIC-1729_Binary_phased_RV}
\end{figure}

\begin{figure}
	\centering
	\includegraphics[width=\textwidth]{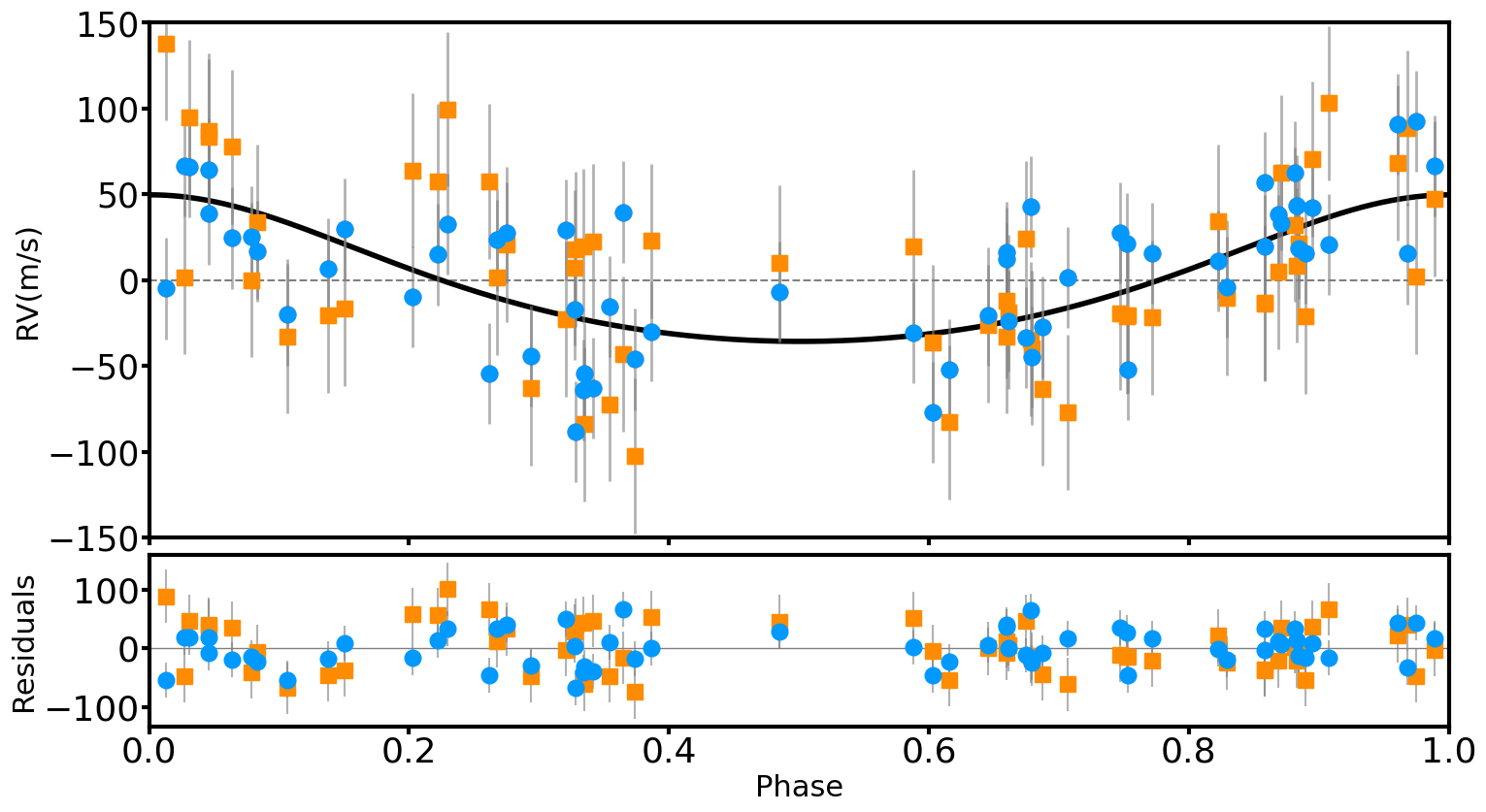}
	\caption{Phased radial velocity plot of highest likelihood \texttt{kima} sample showing the radial velocity motion of planet TIC-1729\,b along with its residuals. The black line is the highest likelihood Keplerian RV signal fit to the data. RV data from the primary star is represented by the blue circles and the secondary star B by the orange squares. The RV data is from \citet{Sairam2024b} obtained using the SOPHIE spectrograph \citep{Perruchot2008}.}
	\label{fig:TIC-1729_phased_RV}
\end{figure}

We present an example analysis of a radial velocity dataset for a tight binary system, aimed at detecting signals of circumbinary planets. Our software of choice is \texttt{kima} \citep{Faria2018}.  One of the main advantages of \texttt{kima} is that all parameters are fit simultaneously, and the number of Keplerian signals ($N_{\rm{p}}$) can be fit as a free parameter. Since \texttt{kima} utilizes a Diffusive Nesting Sampling (DNS) algorithm \citep{Brewer2011}, it computes the Bayesian evidence of the model directly, which can be later used for model comparison. Therefore, the number of Keplerian signals present in the data can be inferred directly from a single fit to the data \citep{Standing2022}. \texttt{kima} has also been adapted specifically for searching circumbinary planet signals. The ``known object'' mode allows the user to provide a separate set of priors for a signal known to be in the dataset and then a set of priors to cover the rest of the parameter space required. This is especially handy when fitting circumbinary planet signals since the secondary star imparts a large radial velocity variation onto the primary star of the order of tens of $km\,s^{-1}$. Tight priors can therefore be placed on the binary signal, and the algorithm can simultaneously fit for additional signals in the data at each step \citep{Standing2022}.

In addition to this, \texttt{kima} possesses a \texttt{kima-binaries} model \citep{Baycroft2023}. This model adapts the algorithm to allow for non-static Keplerian signals like that caused by binary orbital precession. The model also includes the options to fit double-lined (SB2) datasets and the other post-Keplerian effects mentioned above in Section~\ref{sec:cbp_problems} \citep{Baycroft2023}. For further details on the implementation of these changes see \cite{Baycroft2023}.

\texttt{kima} is pip installable and the docs can be found \hyperlink{https://www.kima.science/docs/}{here}\footnote{\url{https://www.kima.science/docs/}}. Once installed, to run \texttt{kima} on a dataset you need only run a python file similar to the template below:

\onecolumn
\begin{python}
	#Import modules
	import os
	import numpy as np
	import kima
	from kima import RVData, RVmodel, BINARIESmodel
	from kima.pykima.utils import chdir
	from kima.distributions import Uniform, Gaussian, ModifiedLogUniform, LogUniform, Kumaraswamy
	
	__all__ = ['TIC-1729']
	
	here = os.path.dirname(__file__) # cwd
	
	#Load in the datafile (for standard targets: double_lined=False)
	data_sophie = RVData('TIC-1729_rvs.rdb', units='kms', skip=2, double_lined=True)
	
	def TIC1729(run=False, **kwargs):
	"""
	Create (and optionally run) an RV model for analysis of TIC-1729 SB2 binary data.
	This loads SOPHIE data from `TIC-1729_rvs.rdb` and creates a model where
	the number of Keplerians is free from 0 to 3.
	
	Args:
	run (bool): whether to run the model
	**kwargs: keyword arguments passed directly to `kima.run`
	"""
	
	data = data_sophie
	#Set the required model
	model = kima.BINARIESmodel(fix=False, npmax=3, data=data)
	#For standard targets use the following line instead of the one above:
	#model = kima.RVmodel(fix=False, npmax=3, data=data) 
	
	
	#Set the priors on the model
	model.Cprior = Uniform(25000,27000) #Systemic velocity (m/s)
	model.Jprior = ModifiedLogUniform(0.1,100) #Jitter (m/s) (added in quadrature to the RV uncertainties)
	
	model.conditional.Pprior = LogUniform(80, 1000) #Period prior (days), inner limit set by binary instability limit)
	model.conditional.Kprior = ModifiedLogUniform(1,200) #RV Semi-amplitude (m/s)
	model.conditional.eprior = Kumaraswamy(0.867,3.03) #Eccentricity (Kumaraswamy prior acts as beta distribution see Standing et al. (2022))
	model.conditional.wprior = Uniform(0,2*np.pi) #Argument of periastron
	model.conditional.phiprior = Uniform(0,2*np.pi) #Starting phase of orbit
	
	#Known object priors (used to set tight priors on the binary) Note: the known object can also be used for any previously detected planet (i.e. transiting)
	model.KO_Pprior = [Gaussian(19.658,0.5)] #Binary period (days)
	model.KO_Kprior = [Gaussian(58549,100)] #Binary semi-amplitude (m/s)
	model.KO_eprior = [Gaussian(0.448,0.05)] #Binary eccentricity
	model.KO_wprior = [Uniform(0,2*np.pi)] #Binary argument of periastron
	model.KO_wdotprior = [Gaussian(0,1000)] #Binary orbital precession (arcsec/yr)
	model.KO_phiprior = [Uniform(0,2*np.pi)] #Binary starting phase of orbit
	model.KO_qprior = [Gaussian(0.972,0.2)] #Binary mass ratio
	
	#Set the sampler parameters (change steps to set length of the run, change threads depending on machine, others are reasonable to leave)
	kwargs.setdefault('steps', 100000) #Number of steps to run
	kwargs.setdefault('num_threads', 12) #Number of threads (depends on number of cores available)
	kwargs.setdefault('num_particles', 3) #Number of MCMC particles to explore parameter space
	kwargs.setdefault('new_level_interval', 25000) #Sample steps taken before a new level created
	kwargs.setdefault('save_interval', 5000) #Steps taken before a sample is saved to the output  
	kwargs.setdefault('thread_steps', 50) #Number of steps each thread takes before communicating level info to other threads
	
	#Model settings only required for binary stars:
	model.star_mass = 1.2368
	model.relativistic_correction = True #Set the model to calculate Post-Keplerian effects
	model.double_lined=True #True if double lined (SB2) binary
	model.eclipsing=True #Set this to True if fitting for an eclipsing binary (inc is then fixed to 90 deg)
	
	if run:
	with chdir(here):
	kima.run(model, **kwargs)
	
	return model
	
	if __name__ == '__main__':
	model = TIC1729(run=True, steps=100000)
	res = kima.load_results(model, diagnostic=True)
	
\end{python}

This example will run \texttt{kima} on the TIC-1729 RV data from \citet{Sairam2024a} when pointed to the RV datafile, using priors informed by the parameters in \citet{Sairam2024a}, and return posterior samples that fit the data. The lines needed to fit circumbinary planets are highlighted in the commented lines, but the same example is easy to adapt for non-circumbinary planets by commenting out these lines and changing the priors for different datasets. Editing the number of steps adjusts the length of the run and the number of posterior samples obtained. We suggest a minimum of 10,000 posterior samples to obtain reliable results. For more information, please see the \texttt{kima} docs.

Figure~\ref{fig:TIC-1729_Np_histogram} demonstrates \texttt{kima}'s ability to determine the number of planetary signals present in a dataset via Bayesian model comparison. The chosen number of signals present in the TIC-1729 dataset is 1 with a Bayes Factor (BF) of $>$1005. The Bayes Factor is the ratio of evidence in favor of one model over another. To calculate this for a \texttt{kima} output, we must simply take the ratio of the corresponding number of posterior samples. In our example, we have obtained 4123 samples with 2 signals and 1005 with 1 signal. Therefore, the BF in favor of 2 planets over 1 planet is then simply $4123/1005=4.1$, which corresponds to weak evidence in favor of the 2-signal model according to the formalism in \citet{Jeffreys1961}. The chosen number of signals is the ratio with the highest value. Since we have obtained 0 samples with 0 signals, the ratio of number of samples with 1 signal over those with 0 signals yields the highest Bayes Factor $1005/0=\infty$. Since an infinite BF is not physical, we approximate that the BF must be $\gg$1005 which passes the BF=150 limit typically used to claim a detection \citep{Jeffreys1961, Trotta2008, Standing2022}. To summarize, we find strong evidence of 1 signal present in the data, while a more complex model with two signals is not statistically favored.

Plotting a histogram of all the orbital periods in all the posterior samples gives Fig.~\ref{fig:TIC-1729_periodogram}. The orbital period of TIC-1729~b from \citet{Sairam2024a} is indicated by the vertical dashed orange line. There is a clear peak in the number of posterior samples with periods corresponding to the planet signal. Plotting the signal from the parameters contained within the posterior sample with the highest posterior likelihood from the samples containing the chosen number of planetary signals gives Figs.~\ref{fig:TIC-1729_Binary_phased_RV} and \ref{fig:TIC-1729_phased_RV}.

\subsection{Confirmation and characterization of transiting planets (TOI-776)} \label{sec:joint_fit}

The transit (Chapter~3) and radial velocity techniques are highly complementary. They both have high sensitivity to short-period planets and small planets if they orbit low-mass stars. Most importantly, they jointly constrain arguably the two most defining properties of a planet: size and mass. The first transiting exoplanet, HD~209458~b \citep{Henry2000ApJ...529L..41H,Charbonneau2000ApJ...529L..45C}, was previously known from radial velocities \citep{Mazeh2000}. In this example, we will do the opposite analysis of HD~209458~b and confirm with radial velocities the discovery of a transiting multi-planet system, TOI-776. 

We use the data presented in \citet{Luque2021A&A...645A..41L} to confirm the existence of two transiting sub-Neptunes orbiting the M~dwarf TOI-776 and measure their masses using radial velocity observations from HARPS. We use \texttt{juliet} \citep{juliet} to jointly fit the transit photometry from TESS and the radial velocities from HARPS. This code written in \texttt{python} is built on several publicly available tools that model transits \citep[\texttt{batman},][]{batman} and radial velocities \citep[\texttt{radvel},][]{radvel}, and it is frequently used in the literature. \texttt{juliet} is pip installable and the docs can be found \hyperlink{https://juliet.readthedocs.io/en/latest/index.html}{here}\footnote{\url{https://juliet.readthedocs.io/en/latest/index.html}}. Once installed, to run \texttt{juliet} on the TOI-776 dataset (radial velocities are available in Table B.1 of \citealt{Luque2021A&A...645A..41L}; TESS data is downloaded as part of the example) use the following template:

\begin{python}
import juliet

# First get TESS photometric data:
times_lc, fluxes, fluxes_error  = juliet.utils.get_all_TESS_data('TOI-776',radius='.005 deg')

# RV data should be in a file similar to 'RVS_TOI776.dat' and have three columns:
# time (BJD), rv (m/s), rv_error (m/s)
# Save that file in the same folder where this script is run

# Now let's define the master prior dictionary 
# First define the TRANSIT priors:
priors = {}

# Name of the parameters to be fit:
params = [
        # Planet parameters
        'P_p1','t0_p1','r1_p1','r2_p1','ecc_p1','omega_p1','K_p1',
        'P_p2','t0_p2','r1_p2','r2_p2','ecc_p2','omega_p2','K_p2',
    
        # Instrument parameters
        # TESS
        'q1_TESS','q2_TESS','sigma_w_TESS',
        # HARPS
        'mu_HARPS','sigma_w_HARPS'
        ]

# Distributions:
dists = [
        # Planets
        'normal','normal','uniform','uniform','uniform','uniform','uniform',
        'normal','normal','uniform','uniform','uniform','uniform','uniform',

        # Instruments
        'uniform','uniform','loguniform'
        'uniform','uniform'
        ]

# Hyperparameters
hyperps = [
        # Planets
        [8.24,0.05],[2458571.41,0.01],[0,1],[0,1],[0,1],[-180,180],[0,20],
        [15.65,0.05],[2458572.60,0.01],[0,1],[0,1],[0,1],[-180,180],[0,20],

        # Instruments
        [0,1],[0,1],[0.01,100.0], 
        [-100,100],[0,20]
        ]


# Populate the priors dictionary:
for param, dist, hyperp in zip(params, dists, hyperps):
    priors[param] = {}
    priors[param]['distribution'], priors[param]['hyperparameters'] = dist, hyperp

# RV data is given in a file, so let's just pass the filename to juliet and load the dataset:
dataset = juliet.load(priors=priors, 
                    t_lc = times, y_lc = fluxes, yerr_lc = fluxes_error, 
                    rvfilename='RVS_TOI776.dat', 
                    out_folder = 'toi776_jointfit',
                    verbose=True)

# And now let's fit it!
results = dataset.fit(n_live_points = 500)
\end{python}

Once the fit is run, the last command will generate a \texttt{juliet.fit} object which has several features the user can explore. The most important is the \texttt{juliet.fit.posteriors} dictionary, which contains three important keys: \texttt{posterior\_samples}, which is a dictionary having the posterior samples for all the fitted parameters, \texttt{lnZ}, which has the log-evidence for the current fit and \texttt{lnZerr} which has the error on the log-evidence. This same dictionary is also automatically saved to the output folder as a \texttt{pickle} file. In addition, a file called \texttt{posteriors.dat} is also printed out, which contains four columns: 1) the parameter name, 2) the median, 3) the upper 68\% credibility band, and 4) the 68\% lower credibility band of the parameter, as extracted from the posterior distribution. The \texttt{juliet} tutorials in the code's webpage can guide the user to produce more outputs, plots, use Gaussian Processes, and much more.

The priors dictionary shows the parameters that are being fit in this model. Five of the planet parameters can be constrained from radial velocity data alone: period ($P$), eccentricity ($e$), argument of periastron ($\omega$), radial velocity semi-amplitude ($K$), and transit mid-time ($t_0$). The latter parameter is used rather than the mean anomaly ($\nu$) because the time of mid-transit ($t_0$) is very precisely constrained by transit observations. The transformation between them is trivial. Two additional parameters can only be constrained from transit ($r_1$ and $r_2$). These two parameters are a reparametrization of the planet-to-star radius ratio ($p \equiv R_p/R_\star$, which relates with the transit depth $\delta$ as $\delta=p^2$) and impact parameter of the orbit ($b$) proposed by \citet{Espinoza2018} to allow only physically plausible values for $p$ and $b$. We define these 7 parameters for each planet in the system. The rest of the parameters are related to the instrument: limb darkening for TESS ($q_1$, $q_2$), an instrumental offset for HARPS ($\mu$), and an additional jitter (white noise) term added in quadrature to the errorbars of TESS and HARPS ($\sigma_{\rm w}$). The priors for each parameter are all uninformative except for the period and mid-time transit which have normal priors centered on their values reported in \citet{Luque2021A&A...645A..41L} to speed up the computation.

A joint model of a planetary system detected in both transit and radial velocity observations allows us to measure the inclination of the system ($i$ from the transit), breaking the mass-inclination degeneracy of the radial velocity technique, and in turn allowing us to calculate the true mass of the planet ($M_{\rm p}$ instead of $M_{\rm p} \sin i$). With the mass and the size, we can infer the planet's bulk density, its surface gravity, and escape velocity. None of these parameters are available if using one of the techniques alone. But, not all transiting planets are realistically amenable for radial velocity follow-up. As they have to be observed with ground-based instruments, factors such as the brightness of the host and its spectral type are crucial. Faint stars, stars that rotate rapidly, or stars with little spectral lines may not be amenable for radial velocity follow-up despite hosting transiting exoplanets. It is also important to estimate the radial velocity semi-amplitude that the transiting planet may exhibit. This can be easily estimated from Eq.~\ref{eq:k2} using the inclination ($i \sim 90^\circ$), orbital period of the planet and host mass from the transit data and assuming a planetary mass compatible with the radius of planet (assuming a certain composition). Empirical mass-radius relationships \citep[e.g.,][]{Lissauer2011,ChenKipping2017,Otegi2020,spright} are not only useful to estimate mass from radius (and viceversa), but also to identify demographic trends in the exoplanet population \citep{HatzesRauer2015,Wolfgang2016,Zeng2019PNAS..116.9723Z,LuquePalle2022}. Once the planet's radial velocity semi-amplitude has been estimated, one can predict the number of measurements and the observing cadence necessary to measure the mass of the planet assuming certain internal instrumental precision. Publicly available tools like \texttt{RVFC} \citep{rvfc} or \texttt{gaspery} \citep{gaspery} are designed to make this calculation even in the presence of stellar activity. The radial velocity follow-up of transiting planets --- discovered mostly by Kepler \citep{Borucki2010} and TESS \citep{TESS} --- has played a pivotal role in the past decade to improve our understanding of small exoplanets, their interior composition, formation, evolution, and atmospheric composition \citep[see reviews by][]{Bean2021,Cloutier2024}.

\section*{Table of available spectrographs}\label{sec:tab_spectrographs}

Available spectrographs are listed in Tables~\ref{tab:Spectrographs_North} and \ref{tab:Spectrographs_South}.

\begin{landscape}
	\begin{table}
		\centering
		\tiny
		\setlength{\tabcolsep}{3pt}
		\caption{High-resolution spectrographs designed for precise radial velocity observations located in the northern hemisphere. Those instruments for which is possible to obtain processed radial velocity measurements directly without the need for reducing raw spectra are marked with `Y' in the Archive row.}
		\label{tab:Spectrographs_North}
		\begin{tabular}{@{}l|p{1.1cm}p{1.6cm}p{1.7cm}p{3.5cm}p{1.3cm}p{2.3cm}p{2.9cm}p{2.9cm}p{0.9cm}p{0.9cm}@{}}
			\hline\hline
			Instrument & Tel. aperture [m] & Wav. range [nm] & Max. resolution $\lambda/\Delta\lambda$ & Location & RV precision [m\,s$^{-1}$] & Calibration & Reference & Example use & First light & Archive [Y/N] \\
            
			\hline
			ELODIE      & 1.93  & 385-680          & 42,000        & Alpes-de-Haute-Provence, France & 13      & HCL            & \cite{Baranne1996}     & \cite{Mayor1995} & 1993  & Y \\
			HIRES       & 10.0  & 300-1100         & 85,000        & Hawai`i, USA                    & 1       & I$_2$C \& HCL  & \cite{HIRES}        & \cite{Harada2024}                     & 1993  & N \\
			FIES        & 2.56  & 370-830          & 67,000        & La Palma, Spain                 & 2-5     & HCL            & \cite{Telting2014}     & \cite{Nespral2017}                    & 2001  & N \\
			SOPHIE      & 1.93  & 387-694          & 75,000        & Alpes-de-Haute-Provence, France & 2-3     & HCL \& FP      & \cite{Perruchot2008}   & \cite{Triaud2022}                     & 2006  & Y \\
			SONG        & 1.0   & 440-690          & 112,000       & La Palma, Spain                 & 5-8     & I$_2$C \& HCL  & \cite{Grundahl2007}    & \cite{Grundahl2017}                   & 2008  & Y \\
			HARPS-N     & 3.58  & 383-690          & 115,000       & La Palma, Spain                 & 0.4-0.7 & HCL \& FP      & \cite{Cosentino2012}   & \cite{Dressing2015}                   & 2012  & Y \\
			CARMENES    & 3.5   & 520-960/960-1710 & 94,600/80,400 & Almer\'ia, Spain                & 1-2     & HCL \& FP      & \cite{Quirrenbach2014} & \cite{Trifonov2018}                   & 2015  & Y \\
			MINERVA     & 4x0.7 & 500-630          & 80,000        & Arizona, USA                    & 2-5     & I$_2$C \& HCL  & \cite{MINERVA}         & \cite{Wilson2019}                     & 2017  & N \\
			IRD         & 8.2   & 970-1750         & 70,000        & Hawai`i, USA                    & 2-3     & LFC \& HCL     & \cite{IRD}             & \cite{Gorrini2023}                    & 2018  & N \\
			HPF         & 10.0  & 810-1280         & 53,000        & Texas, USA                      & 2-3     & LFC            & \cite{Mahadevan2012}   & \cite{Stefansson2020}                 & 2019  & N \\
			MAROON-X    & 8.1   & 500-920          & 88,000        & Hawai`i, USA                    & 0.5     & LFC \& FP      & \cite{Seifahrt2018}    & \cite{Trifonov2021}                   & 2019  & N \\
			SPIRou      & 3.6   & 978-2437         & 70,000        & Hawai`i, USA                    & 2-3     & HCL, FP \& LFC & \cite{SPIRou}          & \cite{Cadieux2022}                    & 2019  & N \\
			EXPRES      & 4.3   & 380-680          & 150,000       & Arizona, USA                    & 0.1-0.3 & LFC            & \cite{Jurgenson2016}   & \cite{Cabot2021}                      & 2020  & N \\
			NEID        & 3.5   & 380-930          & 120,000       & Arizona, USA                    & 0.5     & LFC \& FP      & \cite{Schwab2016}      & \cite{Canas2022}                      & 2020  & Y \\
			PARVI       & 5.0   & 1145-1766        & 100,000       & California, USA                 & 4-10    & LFC            & \cite{Cale2023}        & \cite{Cale2023}                       & 2022  & N \\
			KPF         & 10.0  & 445-870          & 98,000        & Hawai`i, USA                    & 0.5     & LFC \& HCL     & \cite{Gibson2016}      & \cite{Lubin2024}                      & 2022  & N \\
            
			\hline
			HARPS3      & 2.5   & 380-690          & 115,000       & La Palma, Spain                 & 0.1     & HCL \& FP      & \cite{Thompson2016}    & -                                     & NYO   & - \\
			MARVEL      & 4x0.8 & 390-920          & 90,000        & La Palma, Spain                 & 1       & HCL \& FP      & \cite{Raskin2020}      & -                                     & NYO   & - \\
			CHORUS      & 10.4  & 410-780          & 110,000       & La Palma, Spain                 & 1       & HCL, FP \& LFC & -                      & -                                     & NYO   & - \\
			HISPEC      & 10.0  & 980-2460         & 100,000       & Hawai`i, USA                    & 0.3     & LFC \& FP      & \cite{Mawet2019}       & -                                     & NYO   & - \\
			\hline
			\multicolumn{11}{l}{{\bf Notes:} HCL = Hollow cathode lamp; I$_2$C - Iodine cell; FP = Fabry-P\'{e}rot, LFC = Laser Frequency Comb, NYO = Not yet operational.}
		\end{tabular}
	\end{table}
\end{landscape}
	
\begin{landscape}
	\begin{table}
		\centering
		\tiny
		\setlength{\tabcolsep}{3pt}
		\caption{High-resolution spectrographs designed for precise radial velocity observations located in the southern hemisphere.}
		\label{tab:Spectrographs_South}
		\begin{tabular}{@{}l|p{1.1cm}p{1.6cm}p{1.7cm}p{3.5cm}p{1.3cm}p{2.3cm}p{2.9cm}p{2.9cm}p{0.9cm}p{0.9cm}@{}}
			\hline\hline
			Instrument & Tel. aperture [m] & Wav. range [nm] & Max. resolution $\lambda/\Delta\lambda$ & Location & RV precision [m\,s$^{-1}$] & Calibration & Reference & Example use & First light & Archive [Y/N] \\
            
			\hline
			CORALIE     & 1.2   & 390-680   & 60,000    & La Silla, Chile      & 3     & HCL \& FP     & \cite{Queloz2000} & \cite{Martin2019}                  & 1998  & Y \\
			FEROS       & 2.2   & 360-920   & 48,000    & La Silla, Chile      & 7-15  & HCL           & \cite{FEROS}      & \cite{Setiawan2008Natur.451...38S} & 1999  & N \\
			HARPS       & 3.6   & 378-691   & 115,000   & La Silla, Chile      & 0.9   & HCL \& FP     & \cite{HARPS}  & \cite{Mayor2011}                   & 2003  & Y \\
			PFS         & 6.5   & 391-784   & 130,000   & Las Campanas, Chile  & 1.0   & I$_2$C \& HCL & \cite{PFS}        & \cite{Teske2021}                   & 2009  & N \\
			CHIRON      & 1.5   & 415-880   & 79,000    & Cerro Tololo, Chile  & 2-5   & I$_2$C \& HCL & \cite{CHIRON}     & \cite{Jones2015}                   & 2012  & N \\
			ESPRESSO    & 8.2   & 378-788   & 190,000   & Paranal, Chile       & 0.3   & FP \& LFC     & \cite{Pepe2021}   & \cite{Faria2022}                   & 2017  & Y \\
			NIRPS       & 3.6   & 971-1854  &  82,000   & La Silla, Chile      & 1.0   & FP \& LFC     & \cite{NIRPS}      & \cite{Artigau2024}                 & 2022  & Y \\
            
			\hline
			ANDES       & 39.3  & 400-1800  & 100,000   & Cerro Armazones, Chile    & 1     & Unknown   & \cite{Marconi2024}        & - & NYO   & - \\
			G-CLEF      & 25.0  & 350-900   & 108,000   & Las Campanas, Chile       & 0.1   & LFC       & \cite{Szentgyorgyi2016}   & - & NYO   & - \\
			2ES         & 2.2   & 370-850   & 120,000   & La Silla, Chile           & 0.1   & FP \& LFC & \cite{Stuermer2024}       & - & NYO   & - \\
			\hline
			\multicolumn{11}{l}{{\bf Notes:} HCL = Hollow cathode lamp; I$_2$C - Iodine cell; FP = Fabry-P\'{e}rot, LFC = Laser Frequency Comb, NYO = Not yet operational.}
		\end{tabular}
	\end{table}
\end{landscape}

\section*{Table of available open source tools}\label{sec:tab_tools}

Available open source tools are listed in Table~\ref{tab:software}.

\begin{landscape}
	\begin{table}
		\centering
		\tiny
		\setlength{\tabcolsep}{3pt}
		\caption{Summary of open-source tools used for the extraction, analysis, and modeling of radial velocity observations to detect and characterize exoplanets. This is a non-exhaustive list.}	\label{tab:software}
		\begin{tabular}{@{}p{2.0cm}|p{2.0cm}p{2.2cm}p{4.8cm}p{12.6cm}@{}}
			\hline
			\hline
			Usage & Software name & Described in & Available at & Description \\
			\hline\hline
			{RV extraction} & \texttt{SERVAL} & \citet{SERVAL} & \url{https://github.com/mzechmeister/serval} & Measure RV and activity indices from a multitude of instruments using the template-matching technique.  \\ \hline
			& \texttt{raccoon} & \citet{Lafarga2020} & \url{https://github.com/mlafarga/raccoon} & Measure RV and activity indices from a multitude of instruments using the cross-correlation technique. Allows building templates from the observations themselves.  \\ \hline
			& \texttt{s-BART} & \citet{sbart} & \url{https://github.com/iastro-pt/sBART} & Semi-Bayesian implementation of the template-matching technique.  \\ \hline
			& \texttt{wobble} & \citet{wobble} & \url{http://github.com/megbedell/wobble} & Data-driven approach to measure RVs. Allows the contention of high S/N stellar spectra and time-dependent telluric spectra.  \\ \hline
			& \texttt{LBL} & \citet{LBL} & \url{https://lbl.exoplanets.ca/} & Measure RV and activity indices from a multitude of instruments using the line-by-line approach.  \\ \hline
			\hline
			{Signal search}  & \texttt{Period04} & \citet{period04} & \url{http://period04.net/} & Search and fit of sinusoidal signals in time series containing gaps. \\ \hline
			& GLS periodogram & \citet{GLS} & \url{https://github.com/mzechmeister/GLS} & Search and fit of sinusoidal signals and Keplerian orbits in time series containing gaps.  \\ \hline
			& Bayesian GLS periodogram & \citet{BGLS} & \url{https://github.com/j-faria/bgls} & GLS in Bayesian formalism.  \\ \hline
			& $\ell$-1 periodogram & \citet{Hara2017} & \url{https://github.com/nathanchara/l1periodogram} & Optimized GLS to find multiple signals at once, minimizing aliasing.  \\ \hline
			\hline
			{Analysis and modeling} & \texttt{RadVel} & \citet{radvel} & \url{https://radvel.readthedocs.io/en/latest/} & Modeling of Keplerian curves using maximum a posteriori optimization and posterior sampling via Markov Chain Monte Carlo. Allows Gaussian Processes and to build custom models and likelihoods. \\ \hline
			& \texttt{juliet} & \citet{juliet} & \url{} & Modeling of RV (and transit) data using nested sampling. Allows Gaussian Processes and linear decorrelation. \\ \hline
			& \texttt{kima} & \citet{Faria2018} & \url{https://www.kima.science/docs/} & Simultaneous modeling of multiple Keplerian curves using diffusive nested sampling. Allows Bayesian model comparison, Gaussian Processes and planets in binary stars. \\ \hline
			& \texttt{exostriker} & \citet{exostriker} & \url{https://github.com/3fon3fonov/exostriker} & Modeling of RV (and transit) data in a Graphical User Interface. Allows for Keplerian curves or full dynamical models that account for planet-planet interactions in multi-planetary systems. Includes periodograms, MCMC and nested sampling, Gaussian Processes, and long-term stability checks for multi-planetary systems.   \\ \hline
			& \texttt{pyaneti} & \citet{pyaneti2019} & \url{https://github.com/oscaribv/pyaneti} & Modeling of Keplerian curves using maximum a posteriori optimization and posterior sampling via Markov Chain Monte Carlo. Allows for multi-dimensional Gaussian Processes and joint fits with transit data. \\ \hline
			& \texttt{pyORBIT} & \citet{Malavolta2016} & \url{https://github.com/LucaMalavolta/PyORBIT} & Modeling of RVs, light curves, activity indices, and transit-time variations in python. \\ \hline
			& \texttt{EXOFASTv2} & \citet{exofastv2} & \url{https://github.com/jdeast/EXOFASTv2} & Modeling of RV, transit and astrometric data in IDL. Allows modeling the stellar properties simultaneously. \\ \hline
			& \texttt{MCMCI} & \citet{MCMCI} & \url{https://github.com/Bonfanti88/MCMCI} & Modeling of RV (and transit) data using Markov Chain Monte Carlo. Allows modeling the stellar properties using isochrone placement. \\ \hline
			& \texttt{DACE} & \citet{DACE} & \url{https://dace.unige.ch/} & Web platform for exoplanet data, including analysis of RV data. Direct access to public archives and uploading your data. Includes periodograms and linear correlation against stellar activity indices. \\ \hline
			\hline
			
		\end{tabular}
	\end{table}
\end{landscape}

\section*{Glossary}
\addcontentsline{toc}{section}{Glossary}

\begin{description}
  \item[APF] Automated Planet Finder.
  \item[AU] Astronomical Unit.
  \item[AURA] Association of Universities for Research in Astronomy.
  \item[BEBOP] Binaries Escorted By Orbiting Planets.
  \item[BF] Bayes Factor.
  \item[CARMENES] Calar Alto high-Resolution search for M dwarfs with Exoearths with Near-infrared and optical Échelle Spectrographs.
  \item[CCD] Charge-Coupled Device.
  \item[CCF] Cross Correlation Function.
  \item[Cnc] Cancri.
  \item[DACE] Data Analysis Center for Exoplanets.
  \item[DNS] Diffusive Nesting Sampling.
  \item[DOLBY] DOuble-Lined BinarY.
  \item[EXPRES] EXtreme PREcision Spectrograph.
  \item[FP] Fabry-P\'erot.
  \item[FTS] Fourier Transform Spectrometer.
  \item[GLS] Generalized Lomb-Scargle.
  \item[GP] Gaussian Process.
  \item[HARPS] High Accuracy Radial velocity Planet Searcher.
  \item[HD] Henry Draper catalogue.
  \item[HIRES] High Resolution Echelle Spectrometer.
  \item[HRS] High Resolution Spectrograph.
  \item[L-BFGS-B] Limited-memory Broyden–Fletcher–Goldfarb–Shanno Bound algorithm.
  \item[LFC] Laser Frequency Comb.
  \item[MAROON-X] M-dwarf Advanced Radial velocity Observer Of Neighboring eXoplanets.
  \item[MCMC] Markov Chain Monte Carlo.
  \item[NASA] National Aeronautics and Space Administration.
  \item[NIRPS] Near Infra-Red Planet Searcher.
  \item[NOAO] National Optical Astronomy Observatory.
  \item[NSF] National Science Foundation.
  \item[NSO] National Solar Observatory.
  \item[RMS] Root Mean Squared.
  \item[RV] Radial Velocity.
  \item[SB1] Single-lined Spectroscopic Binaries.
  \item[SB2] Double-lined Spectroscopic Binaries.
  \item[SEIS] Second Earth Initiative Spectrograph.
  \item[SERVAL] SpEctrum Radial Velocity AnaLyser.
  \item[S/N] Signal-to-Noise Ratio.
  \item[SOPHIE] Spectrograph for the Observation of the PHenomena of stellar Interiors and Exoplanets.
  \item[TATOOINE] The Attempt to Observe Outer-planets In Non-single-stellar Environments.
  \item[TESS] Transiting Exoplanet Survey Satellite.
  \item[THE] Terra Hunting Experiment.
  \item[TIC] TESS Input Catalogue.
  \item[TOI] TESS Object of Interest.
\end{description}

\bibliography{RV}

\appendix
\section{Supplementary/Online material}\label{chap:appendix_rv}

\subsection{Telescope proposal example}\label{sec:proposal_example}

We provide an example of a successful proposal requesting time on the Very Large Telescope (VLT) with ESPRESSO \citep{Pepe2021} to observe the binary TOI-1338. With previous data, our team was able to identify a candidate planet signal in the observations but were not able to see any trace of the known transiting planet TOI-1338~b \citep{Kostov2020}. The proposal requested time to derive precise radial velocities of the binary to confirm the candidate and transiting planets, and in the process measure their masses. The example below was submitted to the European Southern Observatory (ESO) call for proposals in period P106. It was accepted, and the time was used to gather the data published in \citet{Standing2023}.

More generally, the basic structure of an observing proposal is as follows:
\begin{enumerate}
	
	\item \underline{Background information on the field of research}\\
	Here, try to provide as much information as is needed to explain the current state of the field and set the scene for your proposed target(s). As proposals are typically short, avoid bombarding the reviewing panel with unnecessary information. Try to tell a story and keep the background relevant and brief. 
	
	\item \underline{Information on the target(s) of the proposal}\\
	Introduce the target(s) you are asking to observe. Is it suitable for RV observations? Is there any sign of stellar activity that could impede your proposed results? Add any other noteworthy information about your target, especially that related to previous observations if applicable.
	
	\item \underline{Why is it relevant to observe your target(s)?}\\
	Here it is nice to provide almost a conclusion of sorts, the cliff-notes. Summarize the scientific problem that you aim to address with these observations and how they contribute to advancing the exoplanet field at large.
	
	\item \underline{Observing plan}\\
	What is your goal? How many observations are you requesting? What exposure times do you need for each and why? How often do you want to observe? Are there any constrains related to the weather conditions?
	
	\item \underline{Simulations}\\
	It is always a good idea to provide some simulations to show that you can achieve the results you want with the data you are requesting. Provide figures to show this, and refer to them here.
	
	\item \underline{Data analysis plan}\\
	What precision are you expecting to achieve with your observation? Have you consulted the Exposure Time Calculator (ETC) for the requested instrument? (if it has one). Make sure you or the observatory have the necessary tools to reduce the data and derive the observable you are aiming for (in the example, precise radial velocities of a binary star).
	
	\item \underline{Ancillary science}\\
	This section is typically not compulsory, but it shows that you have thought about how to maximize the science output from the data you request in your proposal. 
	
	\item \underline{Figures and references}\\
	Only include figures that directly support your science case, anything extra is wasting space. Make sure the axes are correctly labeled and that the message from your plot is easy to grasp visually without the need to read the caption or the full proposal. \textbf{Tip:} After creating your plots, arrange them into a single image/figure in an image editor and include the figure caption with a text box. Trim as much excess whitespace as possible, save this figure as an image, and import the image (including the caption) into your proposal. This will allow you to save as much space as possible. For the references, make sure all necessary information is included to find them (i.e., at least one or two authors, journal, volume, and page number).
	
\end{enumerate}

Space is limited in observing proposals. Thus, include only the necessary information to get your points across, and try to be concise while convincing. The Time Allocation Committee (TAC) read hundreds of these in each call for proposals, if you can make yours easy to read, and stand out, you are already at an advantage. \textbf{Tip:} Use bold text to guide the readers to key points.
The TAC normally consists of experts from many different sub-disciplines, therefore you should not assume they will have detailed knowledge of your specific science case.

\textbf{Note:} Some calls for proposals will give more space than others, each with their own set of rules and sections. Be sure to \textit{thoroughly} check these guidelines in the call for proposals before the deadline. Many proposals nowadays are anonymous and require writing in such a way as to hide who the proposing team is. This aids reviewers to focus directly on the science, rather than the scientist. Be careful of this rule if it applies in your case, as revealing the proposer's identity (even unknowingly) can result in automatic disqualification.

\includepdf[pages=-]{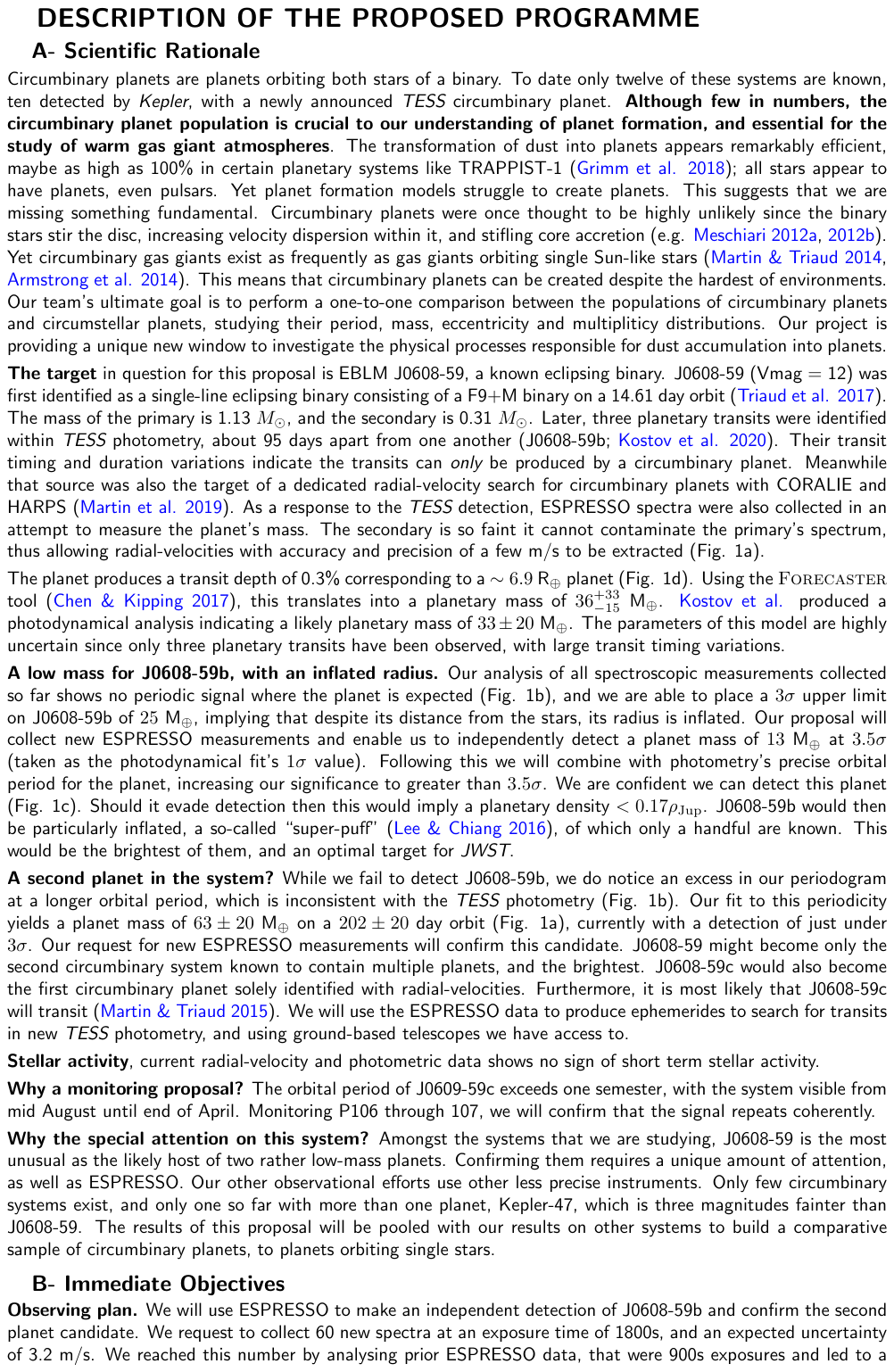} 

\end{document}